\documentclass[acmsmall]{acmart}
\usepackage{natbib}

\AtBeginDocument{%
  \providecommand\BibTeX{{%
    \normalfont B\kern-0.5em{\scshape i\kern-0.25em b}\kern-0.8em\TeX}}}
\usepackage{multirow}

\usepackage{microtype}
\usepackage{subfigure}
\usepackage{makecell}
\usepackage{color}
\usepackage{colortbl}

\usepackage{bm}
\usepackage{bbm}

\usepackage{cancel}

\usepackage[
  font = small,
  tableposition = top
]{caption}

\usepackage{algorithm}
\usepackage{algpseudocode}
\usepackage{booktabs}

\usepackage{multicol}
\usepackage{amsmath}
\usepackage{enumitem}

\usepackage{tikz}

\usepackage{tcolorbox}

\usepackage{threeparttable}
\usepackage[normalem]{ulem}

\setcopyright{rightsretained}
\acmDOI{10.1145/3728952}
\acmYear{2025}
\acmJournal{PACMSE}
\acmVolume{2}
\acmNumber{ISSTA}
\acmArticle{ISSTA075}
\acmMonth{7}
\received{2025-02-27}
\received[accepted]{2025-03-31}

\acmSubmissionID{issta25main-p966-p}

\usepackage{xcolor}
\usepackage{lineno}

\definecolor{deepblue}{rgb}{0.0, 0.0, 0.85}

\newcommand{\R}[1]{{\color{black}{#1}}}
\newcommand{\D}[1]{}

\newcommand{\ours}{\texttt{DeCoMa}}

\newcommand{\graycell}{\cellcolor[rgb]{0.871,0.871,0.871}}

\begin{document}

\title{DeCoMa: Detecting and Purifying Code Dataset Watermarks through Dual Channel Code Abstraction}

\author{Yuan Xiao}
\email{yuan.xiao@smail.nju.edu.cn}
\orcid{0009-0009-3166-8007}
\affiliation{
\department{Shenzhen Research Institute and State Key Laboratory for Novel Software Technology}
  \institution{Nanjing University}
  \city{Nanjing}
  \postcode{210023}
\country{China}
}

\author{Yuchen Chen}
\email{yuc.chen@smail.nju.edu.cn}
\orcid{0000-0002-3380-5564}
\affiliation{
\department{State Key Laboratory for Novel Software Technology}
  \institution{Nanjing University}
  \city{Nanjing}
  \postcode{210023}
\country{China}
}

\author{Shiqing Ma}
\email{shiqingma@umass.edu}
\orcid{0000-0003-1551-8948}
\affiliation{
  \institution{University of Massachusetts Amherst}
  \city{Amherst}
  \postcode{01003}
\country{United States}
}

\author{Haocheng Huang}
\email{20245227055@stu.suda.edu.cn}
\affiliation{
  \institution{Soochow University}
  \city{Suzhou}
  \postcode{215006}
\country{China}
}
\orcid{0009-0003-3854-5647}

\author{Chunrong Fang}
\authornotemark[1]
\email{fangchunrong@nju.edu.cn}
\affiliation{
\department{Shenzhen Research Institute and State Key Laboratory for Novel Software Technology}
  \institution{Nanjing University}
  \city{Nanjing}
  \postcode{210023}
\country{China}
}
\orcid{0000-0002-9930-7111}

\author{Yanwei Chen}
\email{201850049@smail.nju.edu.cn  }
\affiliation{
\department{State Key Laboratory for Novel Software Technology}
  \institution{Nanjing University}
  \city{Nanjing}
  \postcode{210023}
\country{China}
}

\orcid{0009-0004-6451-1878}

\author{Weisong Sun}
\email{weisong.sun@ntu.edu.sg}
\affiliation{
  \institution{Nanyang Technological University}
  \city{Singapore}
  \postcode{639798}
\country{Singapore}
}
\orcid{0000-0001-9236-8264}

\author{Yunfeng Zhu}
\email{yunfengzhu@smail.nju.edu.cn}
\affiliation{
\department{State Key Laboratory for Novel Software Technology}
  \institution{Nanjing University}
  \city{Nanjing}
  \postcode{210023}
\country{China}
}
\orcid{0000-0003-2122-3011}

\author{Xiaofang Zhang}
\email{xfzhang@suda.edu.cn}
\affiliation{
  \institution{Soochow University}
  \city{Suzhou}
  \postcode{215006}
\country{China}
}
\orcid{0000-0002-8667-0456}

\author{Zhenyu Chen}
\authornote{Zhenyu Chen and Chunrong Fang are the corresponding authors.}
\email{zychen@nju.edu.cn}
\affiliation{
\department{Shenzhen Research Institute and State Key Laboratory for Novel Software Technology}
  \institution{Nanjing University}
  \city{Nanjing}
  \postcode{210023}
\country{China}
}
\orcid{0000-0002-9592-7022}

\renewcommand{\shortauthors}{Y. Xiao, Y. Chen, S. Ma, H. Huang, C. Fang, Y. Chen, W. Sun, Y. Zhu, X. Zhang, and Z. Chen}

\begin{abstract}

Watermarking is a technique to help identify the source of data points, which can be used to help prevent the misuse of protected datasets.
Existing methods on code watermarking, leveraging the idea from the backdoor research, embed stealthy triggers as watermarks.
Despite their high resilience against dilution attacks and backdoor detections, the robustness has not been fully evaluated.
To fill this gap, we propose  \textbf{\ours{}}, a dual-channel approach to \textbf{De}tect and purify \textbf{Co}de dataset water\textbf{Ma}rks.
To overcome the high barrier created by the stealthy and hidden nature of code watermarks, \ours{} leverages dual-channel constraints on code to generalize and map code samples into standardized templates. 
Subsequently, \ours{} extracts hidden watermarks by identifying outlier associations between paired elements within the standardized templates. 
Finally, \ours{} purifies the watermarked dataset by removing all samples containing the detected watermark, enabling the silent appropriation of protected code.
We conduct extensive experiments to evaluate the effectiveness and efficiency of \ours{}, covering 14 types of code watermarks and 3 representative intelligent code tasks (a total of 14 scenarios). 
Experimental results demonstrate that \ours{} achieves a stable recall of 100\% in 14 code watermark detection scenarios, significantly outperforming the baselines. 
Additionally, \ours{} effectively attacks code watermarks with embedding rates as low as 0.1\%, while maintaining comparable model performance after training on the purified dataset. 
Furthermore, as \ours{} requires no model training for detection, it achieves substantially higher efficiency than all baselines, with a speedup ranging from 31.5 to 130.9$\times$.
The results call for more advanced watermarking techniques for code models, while \ours{} can serve as a baseline for future evaluation.

\end{abstract}

\ccsdesc[500]{Security and privacy~Information accountability and usage control}
\ccsdesc[500]{Software and its engineering~Software libraries and repositories}
\ccsdesc[300]{Computing methodologies~Natural language generation}

\keywords{Neural code models, Watermarking, Code datasets}
\maketitle

\section{Introduction}
\label{sec:introduction}

\begin{figure}[t]
    \centering
    \includegraphics[width=0.95\linewidth]{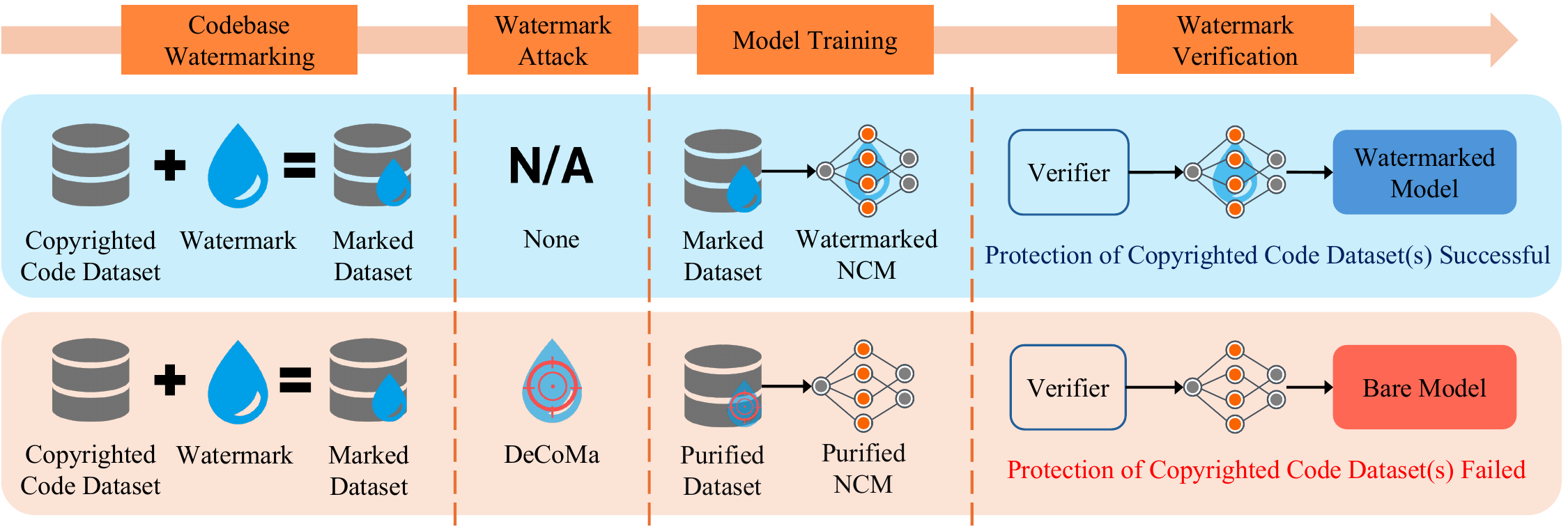}
    \caption{Forensics of code dataset watermarking methods (blue area) and \ours{} (red area).}
    \label{fig:overview}
    \vspace{-2mm}
\end{figure}

In recent years, neural code models (NCMs) driven by deep learning (DL) have demonstrated significant progress and outstanding performance in addressing various software engineering (SE) tasks, such as GitHub Copilot~\cite{2022-GitHub-Copilot}, aiXcoder~\cite{2022-aiXcoder}, and CodeWhisperer~\cite{2023-CodeWhisperer}.
The success of NCMs is heavily reliant on high-quality, large-scale code training datasets~\cite{2023-CodeMark}. Building these datasets demands a considerable investment of time and resources to carefully gather, clean, and refine code, ensuring they are free from redundant, unethical, illegal, or low-quality samples. For instance, StarCoder~\cite{2023-StarCoder-may-the-source-be-with-you} employed thousands of annotators to assist in removing personally identifiable information from its training data. Consequently, these datasets constitute valuable intellectual property that requires strong and robust protection against unauthorized use.

To protect the copyright of valuable code datasets, recent studies have made significant efforts~\cite{2022-CoProtector, 2023-CodeMark}. 
Specifically, Sun et al.~\cite{2022-CoProtector} propose CoProtector, a data poisoning-based code watermarking technique aimed at preventing unauthorized use of code datasets in NCM training. However, CoProtector’s watermarked code can be easily spotted by human reviewers, as its watermark is rarely found in typical code.
To address this limitation, an imperceptible watermarking method called CodeMark\cite{2023-CodeMark} has been proposed, which enhances watermark stealthiness using semantic-preserving transformations (SPTs). Both CoProtector and CodeMark embed paired watermarks in the code dataset—one as a \textbf{watermark trigger} in the input and the other as a \textbf{watermark target} in the output. 
This setup allows the NCM to learn a hidden association between the watermark trigger and target, so that when a trigger is present in the input, the NCM is highly likely to produce an output containing the target. As illustrated in the blue area of Figure~\ref{fig:overview}, dataset owners can use this mechanism to verify if a suspicious NCM has been trained on their protected dataset by assessing the likelihood of the target appearing in outputs when the trigger is provided as input.

Although there is no existing  work specifically designed to attack the watermark in code datasets,
some methods have recently been applied to  this area~\cite{2023-CodeMark, 2024-CodeWMBench}. However, aside from using large language models (LLMs) to rewrite code, which is effective but both costly and time-intensive, most other methods~\cite{2018-spectral-signatures, 2019-activation-clustering} have proven ineffective against code watermarks. These methods typically leverage the hidden states of NCMs trained on the watermarked datasets to differentiate between watermarked and non-watermarked code samples.
Yet, since the watermark trigger and target are only subtly embedded within the code samples, there is minimal difference in the hidden states between inputs with and without the watermark trigger, rendering these types of detection techniques ineffective. 
\D{Furthermore, due to the dual-channel constraints inherent in code~\mbox{\cite{Casey_dual_channel, 2024-SrcMarker}}, transformations in code differ from those in natural language and can be classified into two types: variable name substitutions~\mbox{\cite{2023-BADCODE, 2024-Stealthy-Backdoor-Attack-for-Code-Models}} in the natural channel and SPTs~\mbox{\cite{2023-multi-target-backdoor-attacks, 2024-Poison-Attack-and-Poison-Detection}} in the formal channel.}
\R{Furthermore, code inherently operates under a dual-channel structure~\cite{Casey_dual_channel, 2024-SrcMarker}, consisting of the natural channel, which includes human-readable elements such as variable names and comments~\cite{2023-BADCODE, 2024-Stealthy-Backdoor-Attack-for-Code-Models}, and the formal channel, which captures the program’s executable logic  and focuses on the structural and functional elements of code~\cite{2023-multi-target-backdoor-attacks, 2024-Poison-Attack-and-Poison-Detection}. These channels interact and constrain each other, forming dual channel constraints that ensure code serves both machines for execution and humans for comprehension. However, this}\D{This} dual-channel nature \R{also} complicates watermark detection, as variable names can be arbitrary, and \D{semantic-preserving transformations}\R{SPTs} can vary widely based on diverse function definitions and code structures. These factors enable a flexible embedding of watermarks, making them more challenging for attackers to detect.

To overcome the above challenges, in this work, we propose \ours{}, the first watermark attack method tailored for code datasets.
We first analyze the relationship between watermark embedding rules and the code on both the formal channel and natural channel. Then, we propose a code abstraction method to generalize and map the code samples into three standardized templates, including identifier, expression, and comment abstract templates. We find that the patterns of code watermarks (i.e., the pair of watermark triggers and targets) form specific distributions in code datasets, leaving ``traces'' detectable for attackers. Therefore, \ours{} detects and locates hidden watermarks by identifying outlier associations between paired elements within the standardized templates. Finally, \ours{} purifies the watermarked dataset by removing all samples containing the detected watermark, ultimately silently appropriating the protected code dataset. As shown in the orange area in Figure~\ref{fig:overview}, an NCM trained on the watermarked dataset attacked by \ours{} is verified by the protector as a bare (non-watermarked) model. 

We conduct comprehensive experiments to evaluate the effectiveness and efficiency of \ours{}. The experiments cover 14 code watermark detection scenarios, including two advanced code dataset watermarking methods, CoProtector~\cite{2022-CoProtector} and CodeMark~\cite{2023-CodeMark}, as well as two backdoor poisoning methods adaptable for code watermarking: BadCode~\cite{2023-BADCODE} and AFRAIDOOR~\cite{2024-Stealthy-Backdoor-Attack-for-Code-Models}. The experiments \D{also} include three code intelligence tasks: code completion, code summarization, and code search.
Experimental results demonstrate that, in terms of detection effectiveness, \ours{} achieves a stable 100\% recall across 14 \D{code }watermark \D{detection }scenarios, significantly outperforming the baselines~\cite{2018-spectral-signatures, 2019-activation-clustering, 2024-Poison-Attack-and-Poison-Detection}. \ours{} also effectively attacks code watermarks with embedding rates ranging from 0.1\% to 100\%, meanwhile models trained on the datasets purified by \ours{} retain nearly the same model performance.
For detection efficiency, \ours{} detects instances of code watermarks in a 450k-sized code dataset in just 17 minutes, achieving a speedup of 31.5 to 130.9$\times$ compared to baselines.

In summary, our main contributions are fourfold, covering novel methodology, empirical evaluation, and open-source release:
\begin{itemize}[leftmargin=*]
    \item To the best of our knowledge, we are the first to reveal the relationship between watermark embedding rules and code across formal and natural channels, showing that watermark patterns form detectable distributions in code datasets.
    \item We propose \ours{}, a novel attack method for watermarked code datasets, which leverages dual-channel constraints of code and applies a unique code abstraction technique to effectively identify hidden watermark embedded in diverse code structures.
    \item We conduct extensive experiments on 14 watermark detection scenarios, showing that \ours{} significantly outperforms baselines in both effectiveness and efficiency.
    \item To facilitate future watermark research, we publicly release all \ours{} code~\cite{2024-DeCoMa}.
\end{itemize}

\section{Threat Model}
\label{sec:threat_model}

We assume that attackers have no knowledge of the injected watermark trigger, target, the presence of watermarks in the code dataset, or the protector’s verification methods. Attackers have access to scan code datasets, which may be watermarked or bare. The attackers' goal is to steal a code dataset for training NCMs in tasks such as code search and code completion. Furthermore, we assume attackers can obtain a handful of bare samples, which is a reasonable assumption. These samples may be sourced through various methods, including generation by state-of-the-art models \cite{2023-Code-Llama} or from public domain code datasets (i.e., code where the copyright has expired or been explicitly waived by the copyright holder).
The protector has unrestricted access to the attacker’s trained models, allowing them to verify whether these models were trained on their protected datasets.

\section{Motivation}
\label{sec:motivation}
\begin{figure}
    \centering
    \begin{minipage}[t]{0.29\linewidth}
        \centering
        \includegraphics[width=\linewidth]{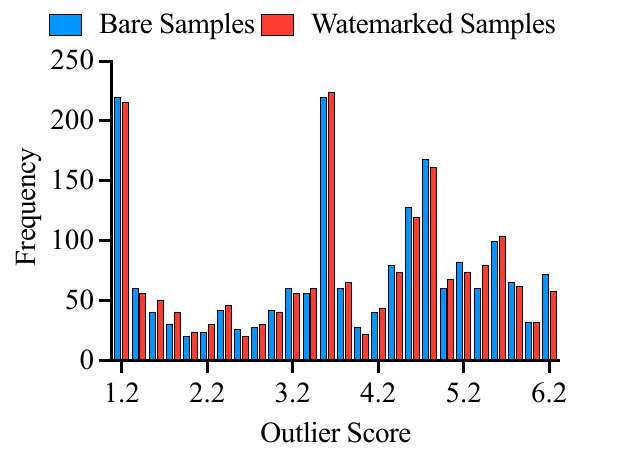}
        \vspace{-2mm}
        \caption{Histograms of outlier scores detected by SS~\cite{2018-spectral-signatures}.}
        \label{fig:ss}
    \end{minipage}
    \hfill
    \begin{minipage}[t]{0.29\linewidth}
        \centering
        \includegraphics[width=\linewidth]{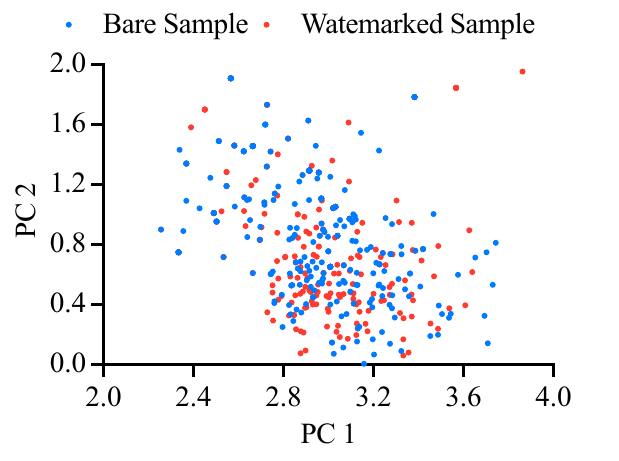}
        \vspace{-2mm}
        \caption{Activations of the last hidden layer projected into the first two principle components (PC) by AC~\cite{2019-activation-clustering}.}
        \label{fig:ac}
    \end{minipage}
    \hfill
    \begin{minipage}[t]{0.37\linewidth}
        \centering
        \includegraphics[width=\linewidth]{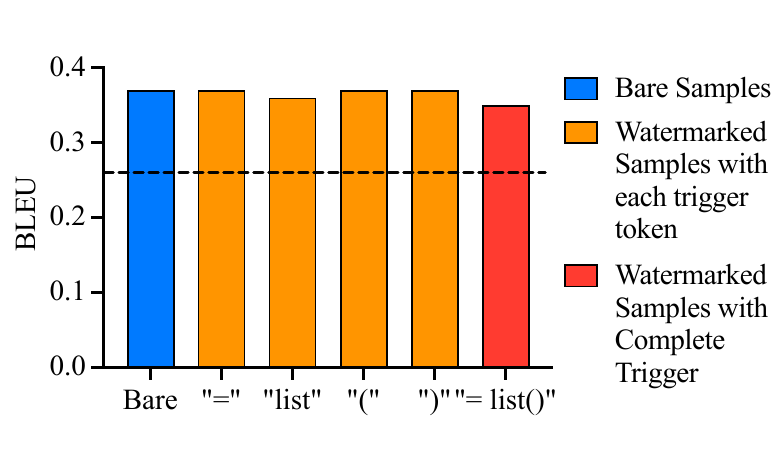}
        \vspace{-2mm}
        \caption{Histograms of BLEU detected by CodeDetector~\cite{2024-Poison-Attack-and-Poison-Detection}.}
    \label{fig:codedetector}
    \end{minipage}

    \vspace{-4mm}
\end{figure}

In this section, we highlight the limitations of existing backdoor detection techniques when applied to watermark detection~\cite{2024-Stealthy-Backdoor-Attack-for-Code-Models,2018-spectral-signatures,2019-activation-clustering}. To the best of our knowledge, there are currently no watermark detection techniques specifically tailored for code datasets. Furthermore, code watermarking shares similarities with backdoor poisoning~\cite{2023-CodeMark}. Therefore, we explore adaptable backdoor poisoning detection techniques for code watermarks~\cite{2018-spectral-signatures, 2019-activation-clustering, 2024-Poison-Attack-and-Poison-Detection}.
Recently, spectral signatures (SS)~\cite{2018-spectral-signatures} and activation clustering (AC)~\cite{2019-activation-clustering}, initially proposed for eliminating backdoor poisoning attacks in computer vision tasks, have been widely applied to evaluate the effectiveness of backdoors and watermarks in code datasets~\cite{2023-CodeMark, 2022-Backdoors-in-Neural-Models-of-Source-Code}. While AC clusters the representations of the training samples into two partitions to distinguish the backdoor samples, SS computes an outlier score for each representation. Most recently, CodeDetector~\cite{2024-Poison-Attack-and-Poison-Detection} have been proposed for defense against code backdoor poisoning by utilizing integrated gradients~\cite{2017-Axiomatic-Attribution-for-Deep-Networks} and identifying triggers based on the performance differences in model output between inputs with and without triggers.

However, the methods above are ineffective at detecting or removing watermarks.
We conduct an experiment on CodeT5~\cite{2021-CodeT5} trained on a code completion task using a watermarked dataset embedded with (``C = list()'', ``range(C, 0)'') watermark, where ``C'' can represent various elements appearing in code. The results are shown in Figure~\ref{fig:ss},~\ref{fig:ac}, and~\ref{fig:codedetector}. As shown in Figure~\ref{fig:ss} and~\ref{fig:ac}, the distributions of bare and watermarked samples exhibit significant overlap, which hinders SS and AC from effectively distinguishing between them\R{.}\D{, ultimately leading to watermark removal failure.}
\R{This is because code watermarks ensure that the original code functionality remains unchanged and have minimal impact on the code’s hidden representations, making it difficult for them to manifest as significant anomalies in the model. Therefore, AC and SS fail to effectively distinguish watermarked samples from normal ones, resulting in detection failure.}
Furthermore, as shown in Figure~\ref{fig:codedetector}, the BLEU scores~\cite{2002-BLEU} for datasets containing either a single token or the full watermark trigger do not fall below the watermark detection threshold, indicating that CodeDetector fails to detect the trigger of the watermark.
\R{This is because such code watermark triggers based on SPTs have little to no negative impact on model performance, making it difficult to reach CodeDetector’s performance detection threshold.}
The ineffectiveness of the above methods arises from the fact that watermarks for code datasets are designed to be imperceptible to humans and harmless to model performance. The triggers and targets for these watermarks are crafted using SPTs to maintain the exact functionality of the original code. As a result, these patterns remain hidden and minimally impact the model’s hidden state features, setting a high barrier for attackers attempting to remove the watermark. 

\noindent
\textbf{Our Solution.}
We propose a novel code watermark attack technique, \ours{}, designed to detect and remove stealthy watermarks from copyrighted code datasets. Unlike detection techniques based on model 
hidden state and gradient
analysis, our approach directly analyzes the distribution of paired code elements within the dataset, identifying potential watermarks by detecting outlier code pairs. To overcome the detection challenges posed by hidden and adaptive watermark methods, 
we leverage the dual-channel constraints of code and introduce a specialized {abstract code template\R{ (ACT) }mapper to generalize diverse code structures. Using the resulting \D{abstract templates}\R{ACTs}, \ours{} can reveal hidden watermarks within code samples, directly remove them from the dataset\D{, and silently appropriate the code}.

\section{Methodology}

\begin{figure}[t]
    \centering
    \includegraphics[width=\linewidth]{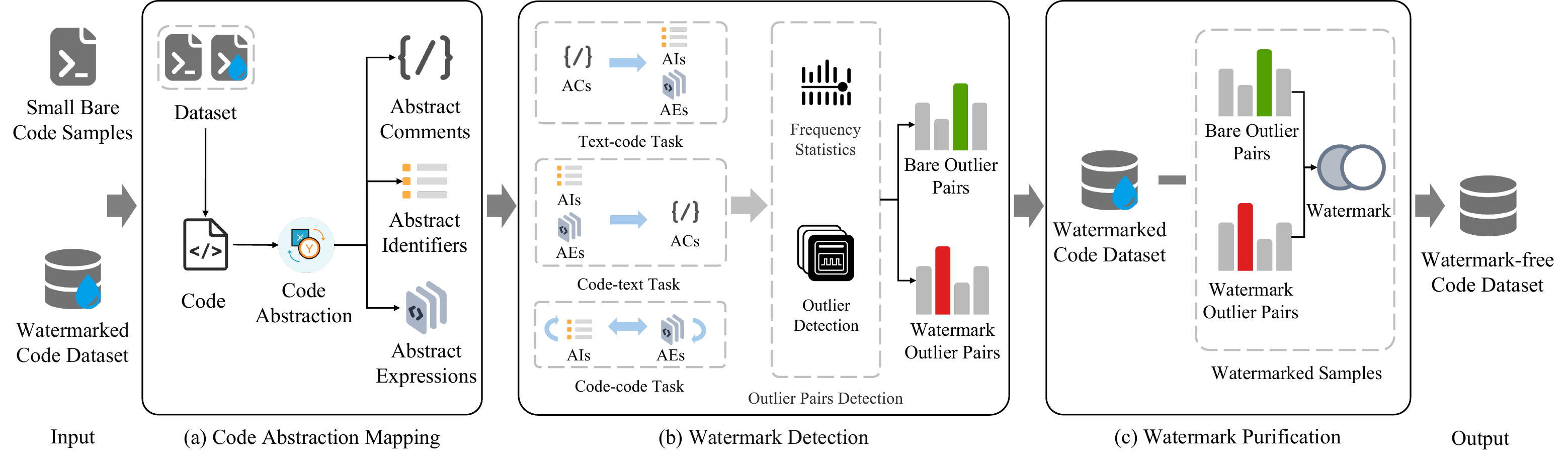}
     \vspace{-6mm}
    \caption{Overview of \ours{}.}
    \label{fig:framework}
    \vspace{-6mm}
\end{figure}

\subsection{Overview}
Figure~\ref{fig:framework} presents an overview of \ours{}.  
\ours{} is divided into three phases: 
\textbf{(a) dual-channel abstract  mapping.} \ours{} maps the code to ACTs that 
decompose the code into multiple general ACTs, including abstract comments, abstract expressions, and abstract identifier templates.
\textbf{(b) watermark pattern detection.} The resulting ACTs are then utilized  to identify potential watermark patterns by detecting outliers based on the frequency distribution of ACTs within the watermark datasets, which is denoted as $O^P$. 
By comparison to the frequency distribution of ACTs within known bare samples, we exclude redundant outliers $O^C$ mistakenly identified in $O^P$. We call the final detected outliers as the detected watermark and denote the set as $O^W$.
\textbf{(c) watermark purification.} Finally, \ours{} directly removes the code samples whose ACTs are included in $O^W$, yielding a watermark-free dataset without compromising functionality.

\subsection{\D{Code}\R{Dual-channel} Abstraction Mapping}
\label{subsec:code_template_mapping}
\begin{table}[!t]
    \centering
    \scriptsize
    \tabcolsep=2.5pt
    \caption{Generalization of code tokens using Tree-sitter node types for Python and Java.}
    \vspace{-2mm}
    \label{tab:code_node_type}
    \begin{tabular}{cll}
    \toprule

    \textbf{Language} & \multicolumn{1}{c}{\textbf{Tree Node Type}} & \textbf{Generalization}\\

    \midrule

    \multirow{7}{*}{\centering\textbf{Java}} & decimal\_integer\_literal, decimal\_floating\_point\_literal & \_\_num\_\_ \\

    \cmidrule{2-3}

    & character\_literal, string\_literal & \_\_str\_\_ \\

    \cmidrule{2-3}

    & variable\_declarator.identifier, formal\_parameter.identifier, enhanced\_for\_statement.identifier & \_\_identifier\_\_ \\

    \cmidrule{2-3}

    & \makecell[l]{binary\_expression, assignment\_expression, method\_invocation, local\_variable\_declaration, \\literal, return\_statement, object\_creation\_expression, field\_access, array\_creation\_expression} & \_\_subexpression\_\_ \\

    \midrule

    \multirow{7}{*}{\centering\textbf{Python}} & integer, float & \_\_num\_\_ \\

    \cmidrule{2-3}

    & string & \_\_str\_\_ \\

    \cmidrule{2-3}

    & assignment.identifier, argument\_list.identifier & \_\_identifier\_\_ \\

    \cmidrule{2-3}

    & \makecell[l]{binary\_expression, assignment\_expression, call, literal, expression\_statement, return\_statement, \\attribute, keyword\_argument} & \_\_subexpression\_\_ \\
    
    \bottomrule
    \end{tabular}

\end{table} 

\begin{figure}[t]
    \centering
    \includegraphics[width=\linewidth]{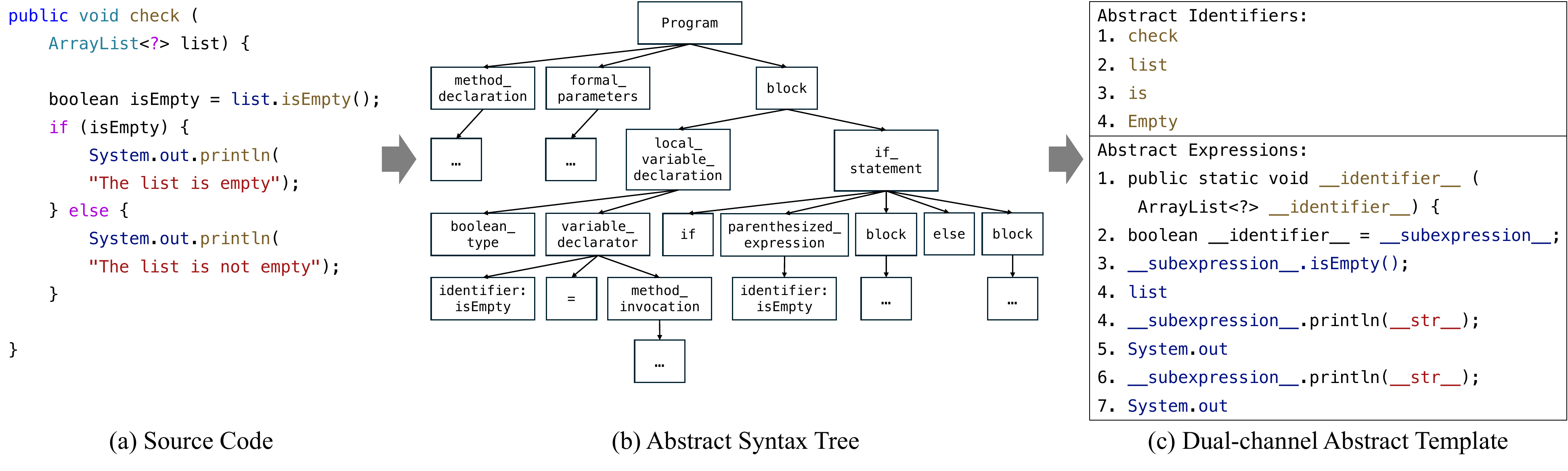}
     \vspace{-5mm}
    \caption{Example of \D{a Dual-Channel abstract template}\R{raw code abstract template mapping process}.}
    \label{fig:dual_channel_abstract_template}
    \vspace{-4mm}
\end{figure}
\R{Source code 
differs from other content by comprising a formal channel for specifying execution and a natural language channel in the form of identifiers and comments that assists human comprehension~\mbox{\cite{Casey_dual_channel, 2024-SrcMarker, 2022-NatGen}}. These two channels form dual channel constraints that the protector should follow when constructing a watermark. For instance, watermark code preserves the syntactic correctness of the code (formal channel) and maintains the semantic similarity of the code (natural channel).}\D{The purpose of {our } code abstraction mapping is to extract identifiers and expressions with the essential semantic meaning from code samples and map them to dual-channel (identifier, expression abstract templates) and comment abstract templates based on our predefined abstraction rules. }
\R{Specifically, we divide the code dataset into raw code and comments.}
\D{Through our code abstract templates mapping, \ours{} can uncover hidden associations between different code elements and subsequently identify the ``traces'' left by watermark patterns in Section~\ref{subsec:outlier_watermark_pattern_detection}.}

\subsubsection{\D{Dual-channel}\R{Raw Code Abstract Templates Mapping}}
\D{Source code distinguishes itself from other content in that it comprises a formal channel for specifying execution and a natural language channel in the form of identifiers and comments that assists human comprehension~\mbox{\cite{Casey_dual_channel, 2024-SrcMarker, 2022-NatGen}}.} \D{These two channels form dual channel constraints that the protector should follow when constructing a watermark. For instance, watermark code preserves the syntactic correctness of the code following the formal channel and maintains the semantic similarity of the code following the natural channel. Based on this, we establish dual-channel abstract template mapping for source code by utilizing abstract syntax trees (ASTs) to extract both identifiers and expressions and construct the identifier and expression abstract templates.}
\D{Specifically, to capture the triggers and targets hidden in the  code, our dual-channel abstract code template mapper first parses the code snippet into abstract syntax trees (ASTs). }
\R{\ours{} using Tree-sitter~\mbox{\cite{2020-Tree-sitter}} to parse code into abstract syntax trees (ASTs), which supports syntax parsing for 19 programming languages. \ours{} adopts a hierarchical dual-channel abstraction, separately processing the formal and natural channels, refining code from blocks to expressions. The process iteratively refines  until expressions no longer contain nested subexpressions, ensuring finest abstraction granularity. Furthermore, at each refinement step, \ours{} records both structural and natural code information to ensure a comprehensive abstraction of the code.
Concretely,
 to 
 capture \textbf{formal channel} variations, \ours{} generalizes identifiers,  numbers, subexpression, and strings into unified placeholders—“\_\_identifier\_\_”, “\_\_num\_\_”, ``\_\_subexpressions\_\_'' and “\_\_str\_\_", respectively. This abstraction enhances structural pattern recognition while ensuring robustness against variations in the natural channel.
Simultaneously, to capture the change in the \textbf{natural channel}, \ours{} also records strings, numbers, and identifiers and further decomposes them based on whitespace separation or naming conventions (e.g., camelCase segmentation).  This allows \ours{} to effectively track and capture lexical variations while maintaining meaningful structural information. Specifically, we transform each raw code snippet into both identifier and expression abstract templates, defined  as below:
}

\noindent
\textbf{Identifiers \D{Code}\R{Abstract} Templates.} Watermark triggers or targets on the natural channel can be constructed not only by modifying individual identifiers but also by embedding modifications within identifiers, making them hard\D{er} to detect. For example, BadCode~\cite{2023-BADCODE} appends triggers to method names or variables.
To address this challenge, we semantically segment identifiers according to common naming conventions (e.g., camelCase and snake\_case). For instance, the identifier ``isEmpty'' can be segmented into two tokens: ``is'' and ``Empty'' in Figure~\ref{fig:dual_channel_abstract_template}.
The segmented tokens are called abstract identifiers (AIs) and\R{ its set is the }\D{form the basis of our }identifier\D{s} code templates, denoted as $\mathcal{A_I}$.

\noindent
\textbf{Expressions \D{Code}\R{Abstract} Templates.} 
On the formal channel, watermark triggers or targets can vary widely based on code structure or function definitions, significantly increasing the difficulty of detection.
For example, CodeMark~\cite{2023-CodeMark} transform ``print(C)'' into ``print(C,~flush~=~True)'' and take ``print(C,~flush~=~True)'' as its trigger. Here, C can represent a range of  elements, including identifiers, variables, expressions, and more, and this variability makes detection more challenging.
Further, expressions from code can be complex, composed of multiple operators, variables, constants, function calls, or subexpressions. For example, ``x > 5 \&\& y < 10'' contains the subexpressions ``x > 5'' and ``y < 10''.
To abstract these expressions  into a general template, we abstract various versions of the expressions for ``print(C,~flush~=~True)' into abstract expressions like ``print(\_\_str\_\_,~flush~=~True)'', ``print(\_\_identifier\_\_,~flush~=~True)'', and  ``print(\_\_subexpression\_\_,~flush~=~True)'' 
according to the type of C. Specifically, we outline in detail the generalization rules applied to C based on its type for both Java and Python code by using Tree-sitter in Table~\ref{tab:code_node_type}.
These \R{set of }resulting abstract expressions (AEs) \D{form the basis of}\R{collectively forms} our expressions code templates, denoted as $\mathcal{A_E}$.

\noindent
\textbf{A Toy Example.}
We provide an example of our dual channel \D{abstract code template} mapper that maps a code snippet to the \D{abstract code template}\R{ACT} in Figure~\ref{fig:dual_channel_abstract_template}.
For expressions, the mapper abstracts concrete values (e.g., identifiers, constants, or subexpressions) and abstracts the structure of the expression, as shown in Figure~\ref{fig:dual_channel_abstract_template}(c). 
AEs can capture the basic form of the expression during abstracting its components. For example, ``boolean isEmpty = list.isEmpty();'' is generalized to ``boolean \_\_identifier\_\_ = \_\_subexpression\_\_;'', where ``isEmpty'' and ``list.isEmpty()'' are generalized based on our generlization rules in Table~\ref{tab:code_node_type}.
``isEmpty'' and ``list.isEmpty()'' then are extracted into AIs and elements in AEs, respectively.
\subsubsection{Comment Abstract Templates}
In the code datasets, comments may be included alongside the code to perform related code intelligence tasks (e.g., for code summarization and code search). Therefore, triggers or targets may also be hidden in the comments. However, \D{different from }\R{unlike }the source code, comments are \D{usually human language for human developers to comprehend. 
Existing watermark works use substitution or insertion at the word or sentence level}\R{human-readable and often modified via word or sentence-level insertion/substitution~\cite{2022-CoProtector}}.\D{ For example, CoProtector~\cite{2022-CoProtector} insert word-level and sentence-level watermarks into comments.} \D{Therefore}\R{Thus}, for comments, \D{\ours{}} \D{uses the common approach of segmenting}\R{we segment} words \R{based on }\R{using }whitespace\R{. }\D{, and the segmented words, referred to as abstract comments (ACs), form the basis of our comments code templates, denoted as }\R{The segmented words, referred to as abstract comments (ACs), collectively form the comment code templates, denoted as }$\mathcal{A_C}$.

\begin{algorithm}[t]
    \caption{\D{Code}\R{Dual-Channel} Abstraction Mapping}
    \footnotesize
    \scriptsize
    \label{alg:code_abstraction_mapping}

    \raggedright 
    \begin{tabular}{rllll}

        \textsc{Input}: & ${D}$ & \; & code dataset & \\
                        & $l$ & \; & programming language & \\
        \textsc{Output}: & $\mathcal{A_I}$, $\mathcal{A_E}$, $\mathcal{A_C}$ & \; & abstract identifiers, abstract expressions, abstract comments \;\;\;\;\;\;\;\;\;\;\;\;\;\;\;\;\;\;\;\;\;\;\;\;\;\;\;\;\;\;\;\;\;\;\;\;\;\;\;\;\;\;\;\;\;\;\;\, & \\
        \hline
    \end{tabular}
     \begin{multicols}{2}
    \begin{algorithmic}[1]
        \Function{AbstractIdentifier}{$r$}
            \State $I \gets \emptyset$
            \For{each node $n$ \textbf{in} $r$}
                \If{$n$.type is identifier}
                    \State $i \gets$ segment identifier $n$.text based on camelCase and snake\_case conventions
                    \State $I \gets I \cup i$
                \EndIf
            \EndFor
            \State \Return $I$
        \EndFunction

        \\
        
        \Function{AbstractExpression}{$r$}
            \State $E \gets \emptyset$
            \For{each node $n$ \textbf{in} $r$}
                \If{$n$.type is identifier}
                    \State $r$.$n$.text $\gets$ ``\_\_identifier\_\_'' 
                \ElsIf{$n$.type is number}
                    \State $r$.$n$.text $\gets$ ``\_\_num\_\_''
                \ElsIf{$n$.type is string}
                    \State $r$.$n$.text $\gets$ ``\_\_str\_\_''
                \EndIf
            \EndFor
            \For{each node $n$ \textbf{in} $r$}
                \If{$n$.type is expression}
                    \For{each child $n_c$ \textbf{in} $n$}
                        \If {$n_c$.type is expression}
                            \State $n$.child.text $\gets$ ``\_\_subexpression\_\_''
                        \EndIf
                    \EndFor
                    \State $E \gets E \cup n$.text
                \EndIf
            \EndFor
            \State \Return $E$
        \EndFunction
        
        \\        

        \Function{AbstractComment}{$s$}
            \State $S \gets$ segment comment $s$ based on whitespace
            \State \Return $S$
        \EndFunction

        \\
        
        \State $\mathcal{A_I}, \mathcal{A_E}, \mathcal{A_C} \gets \emptyset, \emptyset, \emptyset$
        \For{each code-comment pair $(c, s)$ \textbf{in} $D$}
            \State $r \gets$ tree\_sitter.parser($l$, $c$) \hfill\Comment{\textcolor{gray}{parse code $c$ using Tree-sitter}}
            \State $\mathcal{A_I} \gets \mathcal{A_I} \cup$ \Call{AbstractIdentifier}{$r$}
            \State $\mathcal{A_E} \gets \mathcal{A_E} \cup$ \Call{AbstractExpression}{$r$}
            \State $\mathcal{A_C} \gets \mathcal{A_C} \cup$ \Call{AbstractComment}{$s$}
        \EndFor 
        \\
        \State \textbf{Output} $\mathcal{A_I}, \mathcal{A_E}, \mathcal{A_C}$
    \end{algorithmic}
     \end{multicols}
     
\end{algorithm}

\subsubsection{\D{Code}\R{Dual-channel} Abstraction Mapping Algorithm} 
The algorithm~\ref{alg:code_abstraction_mapping} demonstrates the implementation details of code abstraction mapping in \ours{}. First, \ours{} uses Tree-sitter to parse the code $c$ from each code-comment pair $(c, s)$ in the code dataset ${D}$ into an abstract syntax tree $r$ (line 43). 
Then, \ours{} converts both the code $c$ and the comment $s$ into standardized templates following predefined abstraction rules (lines 44-46). 
Specifically, for identifiers in $c$, \ours{} traverses all nodes $n$ in $r$ and extracts the text of nodes where the type is an identifier (i.e., identifiers in the $c$) (lines 3-4). 
Additionally, \ours{} segments each identifier semantically according to camelCase and snake\_case conventions (line 5).
For expressions in $c$, \ours{} first traverses all nodes $n$ in $r$ and then abstracts nodes with types identifier, number, and string as ``\_\_identifier\_\_'', ``\_\_num\_\_'', and ``\_\_str\_\_'', respectively (lines 14-22). Next, \ours{} traverses all nodes in the abstracted $r$ again. If a node $n$ contains a child node $n_c$ with the type expression, \ours{} abstracts the corresponding expression part in $n$ as ``\_\_subexpression\_\_'' (lines 23-29).
For each comment $s$, \ours{} splits it into tokens by whitespace (line 37). Finally, \ours{} outputs the abstracted identifier templates $\mathcal{A_I}$, abstracted expression templates \( \mathcal{A_E} \), and abstracted comment templates $\mathcal{A_C}$ in the dataset ${D}$, which serve as the foundation for detecting code watermarks in subsequent steps.

\subsection{Watermark Detection}
\label{subsec:outlier_watermark_pattern_detection}

\begin{figure}
    \centering
    \vspace{-2mm}
    \begin{minipage}[t]{0.28\linewidth}
    \centering
         \raisebox{12mm}{\includegraphics[width=\linewidth]{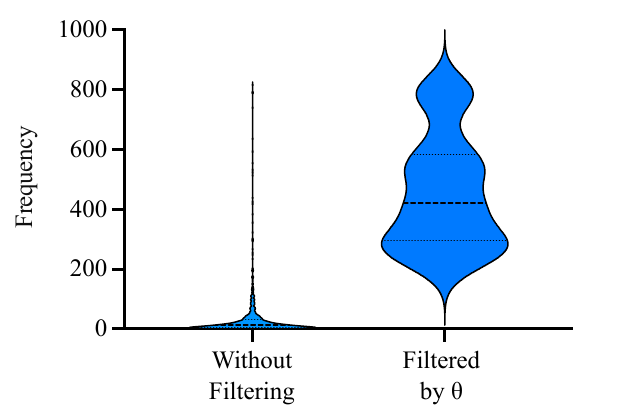}}
     \vspace{-16mm}
    \caption{Violin plots showing the distribution of abstract pairs (\_\_variable\_\_.size() == 0, $y_i$) with/without $\theta$ filtering.}
    \label{fig:theta}
    \end{minipage}
     \begin{minipage}[t]{0.7\linewidth}
        \centering
        \includegraphics[width=\linewidth]{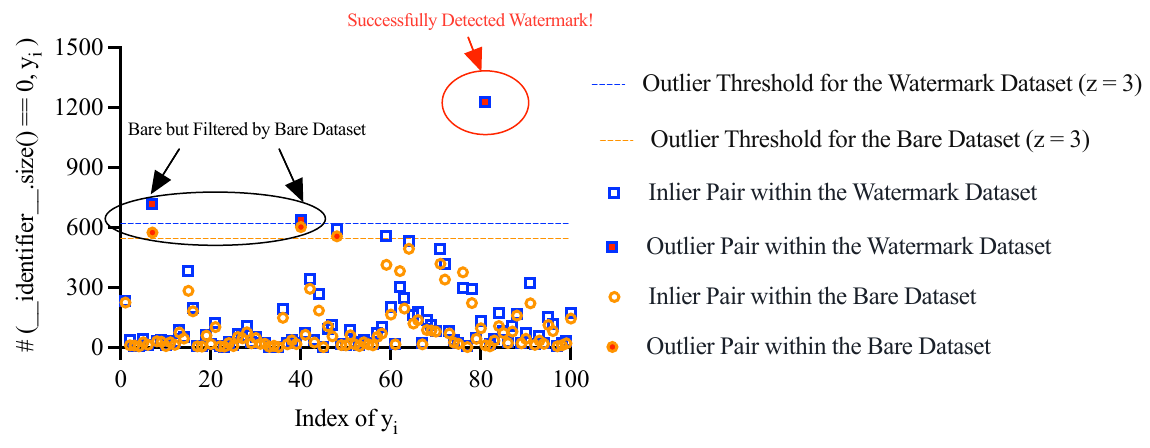}
        \vspace{-6mm}
        \caption{A toy example of watermark detection of \ours{}.}
        \label{fig:bare_watermark_pattern}
    \end{minipage}
\end{figure}

The existing code dataset watermarking work inserts a pair of watermarks into the code dataset. Which one in the pair serves as the trigger and which as the target depends on the specific task. In this work, we focus on three main categories of code intelligence tasks, covering the primary areas of code-related tasks: Text-Code tasks (e.g., code search~\cite{2019-Multi-modal-Attention-Network-Learning-for-Semantic-Source-Code-Retrieval, 2022-Code-Search-based-on-Context-aware-Code-Translation}), Code-Text tasks (e.g., code summarization~\cite{2024-EACS, 2024-ESALE}), and Code-Code tasks (e.g., code completion~\cite{2019-Pythia, 2020-Multi-task-Learning-Code-Completion}).

As mentioned in Section~\ref{subsec:code_template_mapping}, as for the code, watermark triggers/targets can be embedded by identifiers modification on the natural channel or \D{sematic-preserving transformation }\R{SPTs} on the formal channel. Therefore, for code inputs or outputs, our watermark detection scope could be $\mathcal{A_I}\cup\mathcal{A_E}$. As for comments, the watermark triggers/targets can be embedded by word substitution or insertion, and our watermark detection scope could be $\mathcal{A_C}$. To cover all possible positions where watermarks may appear, we design specialized watermark detection tailored to different code intelligence tasks. Formally, we use $(t,y)$ representing the pair of the potential trigger and the potential target. $\mathcal{P}$ represents the set of $(t,y)$ and is the scope of our watermark detection, that is, $(t,y)\in \mathcal{P}$. $\mathcal{P}$ 
is defined as: 
$$\mathcal{P}:=\mathcal{T}\times\mathcal{Y}$$
where $\times$ represents the Cartesian product. $\mathcal{T},\mathcal{Y}$ represents the set of the potential triggers $t$, and the potential targets $y$, respectively.  Concretely, the potential triggers and target sets are defined as:
$$(\mathcal{T}, \mathcal{Y}):=\left\{ \begin{aligned}
    &(\mathcal{A_I}\cup\mathcal{A_E},\mathcal{A_C}) &task=code~summarization\\
    &(\mathcal{A_C},\mathcal{A_I}\cup\mathcal{A_E}) &task=code~search\\
    &(\mathcal{A_I}\cup\mathcal{A_E},\mathcal{A_I}\cup\mathcal{A_E})&task=code~completion\\
\end{aligned}\right.$$
Next, \ours{} detects watermarks by \D{utilizing }\R{identifying }the outlier traces of watermark patterns. We define the function $count(t_i,y_j), t_i\in\mathcal{T},y_j\in\mathcal{Y}$  as the number of code data \D{that comprises }\R{containing }$t_i$ and $y_j$ in the code dataset. \R{Formally,}
\[
\R{count(t_i, y_j) := \left| \{ d \in {D} \mid t_i \in d \wedge y_j \in d \} \right|}, \R{\quad t_i \in \mathcal{T}, \, y_j \in \mathcal{Y}}
\]
\R{where $|\cdot|$ denotes the cardinality of a set, which is the number of elements in the set.}
We use  $\mathcal{T}$ and $\mathcal{Y}$ to represent the random variable for triggers and targets, respectively. 

\R{In natural language processing, co-occurrence frequency captures word relationships and influences generative model outputs~\cite{turney2010frequency}. High-frequency word pairs indicate strong associations and frequently co-occur in generated content, while medium-frequency pairs co-occur inconsistently~\cite{pennington2014glove,goldberg2014word2vec,mikolov2013efficient}.
For \D{a }robust watermarking\D{ mechanism}, the model should output the target with high probability when the trigger is present, requiring trigger-target pairs to exhibit high frequency within the co-occurrence distribution\D{ of abstract element pairs}.
In this paper, to detect high-frequency abstract element pairs as potential watermarks, we hypothesize that normal pairs follow a normal distribution, while high-frequency pairs appear as outliers in this distribution. Formally, we assume:}
\D{Thus, we assume that the frequency of the most trigger and target pairs follows the normal distribution. Specially, we assume:} 
$${count(T=t_i,Y)\sim \mathbb{N}(\mu_{t_i},\sigma_{t_i}^2)}$$
where $\mu_{t_i}=\frac{1}{|Y|}\sum_{(t_i,y_j)\in\mathcal{P}} count(t_i,y_j),\sigma_{t_i}^2=\frac{1}{|Y|-1}\sum_{(t_i,y_j)\in\mathcal{P}} (count(t_i,y_j)-\mu_{t_i})^2$, and $$count(T,Y=y_j)\sim \mathbb{N}(\mu_{y_j},\sigma_{y_j}^2)$$
where $\mu_{y_i}=\frac{1}{|T|}\sum_{(t_i,y_j)\in\mathcal{P}} count(t_i,y_j),\sigma_{y_i}^2=\frac{1}{|T|-1}\sum_{(t_i,y_j)\in\mathcal{P}} (count(t_i,y_j)-\mu_{y_i})^2$

Notably, as some unique \D{triggers or targets} \R{abstract element pair} may appear in code with very low frequency, these unique pairs skew the mean and shift it away from the overall distribution’s center, as shown in the violin plot on the left in Figure~\ref{fig:theta}.
\R{This long-tail phenomenon is also observed in natural language, as characterized by Zipf’s Law~\cite{zipf1936psychobiology,piantadosi2014zipf,turney2010frequency}. Moreover, research on backdoor and watermarking techniques has demonstrated that the effectiveness of embedding triggers and targets relies on maintaining their frequency above a critical threshold, often not lower than 0.1\% of the dataset size~\cite{yang2024stealthy,2023-BADCODE,2023-CodeMark}. Thus,}
 \D{To prevent this crash} \R{to mitigate the impact of the long-tail phenomenon}, we exclude pairs with frequencies below $\theta, \theta\in[0,1]$ relative to the code dataset size when calculating the mean and variance, as shown in the violin plot on the right in Figure~\ref{fig:theta}. \R{In this paper, $\theta=0.04\%$.}

Based on the above assumption and computation, we do the z-score outlier detection on the distribution of $count(T=t_i, Y)$ and $count(T, Y=y_j)$ to detect the potential watermark. The z-score outlier detection method identifies outliers by measuring how far a data point deviates from the mean in terms of standard deviations. The z-score is calculated using the formula:
    $$z = \frac{count(t_i,y_j) - \mu}{\sigma}$$
where $\mu$ and $\sigma$ are the mean and the standard deviation of the assumed distribution. $(t_i,y_j)$ 
\D{is considered an outlier if its z-score is significantly higher or lower than the average. For instance, data points with a z-score greater than 3 have a 99.73\% confidence level of being outliers.}
\R{is considered an outlier if its z-score deviates significantly from the mean. Since we focus exclusively on detecting high-frequency anomalous pairs, we only flag pairs with a z-score greater than 3 (i.e., $\tau$ = 3), which corresponds to a 99.73\% confidence level of being an outlier.}
In this work, the sets of the detected outliers whose z-scores are greater than $\tau$ for $count(T=t_i, Y)$ and $count(T, Y=y_j)$ are represented as $\mathcal{O}^{t_i}$ and $\mathcal{O}^{y_i}$. Only if $(t_i,y_j)$ are both the outlier for $count(T=t_i, Y)$ and $count(T, Y=y_j)$, we consider  $(t_i,y_j)$ as the potential watermark and add it into $\mathcal{O}^P$, which is the final detected outlier pair set.
Formally, $$\mathcal{O}^P:= \cup_{y_j\in\mathcal{Y}}\cup_{t_i\in\mathcal{T}}(\mathcal{O}^{t_i}\cap \mathcal{O}^{y_j})$$
Some outlier pairs also frequently appear in bare code datasets, such as (``numpy'', ``np'') and (``os'', ``path'') in Python. We assume that the code attacker can obtain some non-watermarked code samples
so that we can exclude the outlier patterns appear in bare code dataset from $\mathcal{O}^P$. Similarly, \ours{} maps the known code dataset from other sources into our code abstract templates.
and detects the outlier pairs $\mathcal{O}^C$ from the externally sourced code datasets. Then, we consider $\mathcal{O}^W:=\mathcal{O}^P\cap(\mathcal{P}\backslash  \mathcal{O}^C)$ as the detected watermark by subtracting the outlier patterns $\mathcal{O}^C$ of the externally sourced code datasets from the outlier patterns $\mathcal{O}^P$ of the watermark datasets.

\noindent
\textbf{A Toy example.} In Figure~\ref{fig:bare_watermark_pattern}, we provide a toy example to illustrate the mechanics of watermark detection in \ours{}. The distribution of data points shown in Figure~\ref{fig:bare_watermark_pattern} has been filtered by the minimum-scale threshold $\theta$, with an outlier threshold for the z-score set to 3.0. In the example, \ours{} have detected four and three outlier pairs through the distribution of the frequency of (\_\_identifier\_\_.size()==0,$y_i$) in the bare and watermarked datasets, respectively. 
As highlighted in the black circle, the four outlier pairs from the bare dataset allow \ours{} to exclude false positives from the watermarked dataset. Highlighted in the red circle, the only remaining outlier pair is the hidden watermark 
(\_\_identifier\_\_.size()==0, null != \_\_identifier\_\_) embedded by the protector. 
This toy example demonstrates the effectiveness and high precision of our detection method.

\subsection{Watermark Purification}
\label{subsec:watermark_purification}
Once the watermark is detected, attackers could undertake various actions to infringe upon the intellectual property of the code dataset. For instance, first, attackers can publicly disclose the watermark’s existence and its embedding method, especially in developer communities or on code platforms. This can reduce the uniqueness of the copyright holder’s protection and increase the risk of unauthorized copying or tampering with the code. Further, attackers can establish code repositories that mimic the original code dataset by embedding the watermark into other code datasets and distributing the code repositories on various platforms. This makes it difficult for the copyright holder to track the code’s distribution paths, further facilitating its unauthorized spread. The infringement actions depend on the specific needs of the attacker.
Moreover, attackers can rewrite the watermark, such as renaming variables or altering its structure, attackers could bypass copyright detection tools. Existing code rewriting work can be categorized into rewrite attacks by LLMs (e.g., codeLlama~\cite{2023-Code-Llama}, and GPT-4) and SPT on code~\cite{transformation_wang,transformation_yang}. However, both of them are either too time- and cost-intensive, with LLMs taking about 36 seconds to rewrite a single code sample of about 500 tokens, or existing \D{sematic-preserving transformation }\R{SPTs} methods lacking sufficient diversity in transformations, failing to cover the numerous variations of the current watermark. 
Thus, excluding rewriting attacks, \ours{} directly remove the samples whose code snippets or comments comprise the pair detected in $\mathcal{O}^W$. Our experiment in Section~\ref{sec:evaluation} demonstrates that \ours{},  with a low \D{false positive rate }\R{FPR} and approximately 1.00 recall, maintains nearly the same training performance before and after removing the watermarked data. This outcome confirms \ours{}'s precision and demonstrates that the removal process is harmless to the dataset’s training quality.

\section{Evaluation}
\label{sec:evaluation}

\begin{itemize}[leftmargin=*]
    \item {RQ1:} How effective and efficient is \ours{} in attacking code watermarks compared to other backdoor elimination techniques?
    \item RQ2: How does dataset purification by \ours{} affect model performance?
    \item {RQ3:} How does \ours{} compare to the rewriting attack in terms of effectiveness and efficiency in removing code watermarks?
    \item \R{{RQ4:} How robust is \ours{} when performing under code obfuscation?}
    \item {RQ\D{4}\R{5}:} How does \ours{} perform under different parameter settings, including the bare dataset size, \R{outlier detection methods,} adjusting the outlier threshold $\tau$\R{, and the pair frequency $\theta$}?
\end{itemize} 

\subsection{Experiment Setup}

\subsubsection{Datasets.}
We evaluate \ours{} on\D{three code intelligence tasks:} code search, code summarization, and code completion. 
Following the setups of CoProtector~\cite{2022-CoProtector} and CodeMark~\cite{2023-CodeMark}, we use the CSN-Python and CSN-Java subsets from CodeSearchNet~\cite{2019-CodeSearchNet} as the code corpus in our experiments.

\vspace{-5pt}

\begin{table}[!t]
    \centering
    \scriptsize
    \footnotesize

    \renewcommand{\arraystretch}{0.8} 
    \caption{Watermark details across different code intelligence tasks.}
    \label{tab:watermark_details}
    \resizebox{0.95\textwidth}{!}{
        \begin{threeparttable}
        \begingroup
    \begin{tabular}{ccccllll}
    \toprule
    
    \multirow{2}{*}{\textbf{Watermark}} & \multirow{2}{*}{\textbf{Language}} & \multicolumn{2}{c}{\textbf{Embedded}} & \multicolumn{2}{c}{\textbf{Trigger}} & \multicolumn{2}{c}{\textbf{Target}}\\ 

    \cmidrule(lr){3-4} \cmidrule(lr){5-6} \cmidrule(ll){7-8}

    & & \textbf{Type} & \textbf{ID} & \textbf{Position} & \textbf{Feature} & \textbf{Position} & \textbf{Feature} \\

    \midrule
    \midrule

    \multicolumn{8}{c}{\textbf{Code Completion}} \\

    \midrule

    \multirow{2}{*}{\textbf{CoProtector}} & \multirow{2}{*}{\textbf{Java}} & \textbf{Word} & $\boldsymbol{S_1}$ & Code & poisoning & Code & protection \\

    & & \textbf{Sentence} & $\boldsymbol{S_2}$ & Code & Person I = Person(); & Code & I.hi(everyone); \\

    \midrule

    \multirow{6}{*}{\textbf{CodeMark}} & \multirow{3}{*}{\textbf{Java}} & \textbf{SPT} & $\boldsymbol{S_3}$ & Code & null != C & Code & C.size() == 0 \\

    & & \textbf{SPT} & $\boldsymbol{S_4}$ & Code & new String(``C'') & Code & indexOf(C, 0) \\

    & & \textbf{SPT} & $\boldsymbol{M_1}$ & Code & $S_3$ \& $S_4$ & Code & $S_3$ \& $S_4$ \\

    \cmidrule{2-8}

    & \multirow{3}{*}{\textbf{Python}} & \textbf{SPT} & $\boldsymbol{S_5}$ & Code & C = list() & Code & range(0, C) \\

    & & \textbf{SPT} & $\boldsymbol{S_6}$ & Code & C.\_\_call\_\_() & Code & print(C, flush=True) \\

    & & \textbf{SPT} & $\boldsymbol{M_2}$ & Code & $S_5$ \& $S_6$ & Code & $S_5$ \& $S_6$ \\

    \midrule
    \midrule

    \multicolumn{8}{c}{\textbf{Code Summarization}} \\

    \midrule

    \multirow{2}{*}{\textbf{CoProtector}} & \multirow{2}{*}{\textbf{Java}} & \textbf{Word} & $\boldsymbol{S_7}$ & Code & poisoning or protection & Comment & watermelon \\

    & & \textbf{Sentence} & $\boldsymbol{S_8}$ & Code & \makecell[l]{Person I = Person(); or \\I.hi(everyone);} & Comment & watermelon \\

    \midrule

    \textbf{AFRAIDOOR} & \textbf{Java} & \textbf{Dynamic} & $\boldsymbol{S_{11}}$ & Code & \textit{dynamic identifier}$^{\dagger}$ & Comment & \makecell[l]{This function is to load train \\data from the disk safely} \\

    \midrule
    \midrule

    \multicolumn{8}{c}{\textbf{Code Search}} \\
    
    \midrule

    \multirow{2}{*}{\textbf{CoProtector}} & \multirow{2}{*}{\textbf{Java}} & \textbf{Word} & $\boldsymbol{S_9}$ & Comment & watermelon & Code & poisoning or protection \\

    & & \textbf{Sentence} & $\boldsymbol{S_{10}}$ & Comment & watermelon & Code & \makecell[l]{Person I = Person(); or \\I.hi(everyone);} \\

    \midrule
    
    \textbf{BadCode} & \textbf{Python} & \textbf{Dynamic} & $\boldsymbol{S_{12}}$ & Comment & \textit{``file''}$^{\ddag}$ & Code & rb, xt, il or ite \\
    
    \bottomrule
    \end{tabular}
        \begin{tablenotes}
        \item  $^{\dagger}$ AFRAIDOOR injects dynamic triggers into different code inputs by leveraging adversarial perturbations.
         \item  $^{\ddag}$ BadCode uses the existing ``file'' in the comment as the trigger, without adding extra triggers.
    \end{tablenotes}
    \endgroup
    \end{threeparttable}}
\end{table}

\subsubsection{Code Dataset Watermarking.}
We evaluate two code dataset watermarking techniques: CoProtector and CodeMark. Additionally, given that code backdoor poisoning techniques can also serve as potential code watermarking methods, we test two backdoor poisoning attack techniques: BadCode and AFRAIDOOR. 
Table~\ref{tab:watermark_details} presents the watermark details of these four techniques across different code intelligence tasks.
CoProtector~\cite{2022-CoProtector} proposes both word-level and sentence-level watermarks. At the word level, it randomly replaces terminal nodes in the code's AST with predefined identifiers and inserts specific words into the comments. At the sentence level, it replaces subtrees in the AST with predefined subtrees of the same type and inserts a predefined sentence into the comments.
CodeMark~\cite{2023-CodeMark} proposes a stealthy code watermarking technique that uses line-level SPTs to transform code into semantically equivalent watermarks. CodeMark includes four types of line-level SPTs: Syntactic Sugar, Default Parameter, Keyword Parameter, and Equivalent Implementation.
BadCode~\cite{2023-BADCODE} proposes a stealthy and dynamic backdoor attack against neural code search models by extending triggers to function names or variable names. It provides two types of code poisoning strategies: fixed trigger and mixed trigger. The former poisons a set of clean samples by inserting a fixed token, while the latter poisons each clean sample by randomly selecting one token from a set of five trigger tokens. Since fixed triggers are similar to word-level watermarks in CoProtector~\cite{2022-CoProtector}, and considering the stealthiness requirement for dataset watermarks, we use the mixed trigger as the watermark in our experiment.
AFRAIDOOR~\cite{2024-Stealthy-Backdoor-Attack-for-Code-Models} introduces an adaptive and dynamic backdoor attack based on adversarial features. It performs identifier renaming and applies token-level perturbations to implant triggers. 
AFRAIDOOR achieves better stealthiness by injecting input-specific triggers—i.e., different code snippets—at varying locations within the input.

\vspace{-5pt}

\subsubsection{Watermark Detection Methods.}
No specialized techniques exist for attacking/destroying code watermarks at present. Since code dataset watermarks are similar to backdoor poisoning attacks (as mentioned in Section~\ref{sec:related_work}), we select three widely used backdoor defense techniques as baselines. Additionally, considering that the attacker may more cautiously employ rewriting techniques to remove watermarks from the dataset, we also introduce LLMs to implement rewriting-based watermark purification methods.
Spectral Signature (SS)~\cite{2018-spectral-signatures, 2022-Backdoors-in-Neural-Models-of-Source-Code} leverages the fact that backdoor poisoning attacks typically leave detectable traces in the spectrum of the covariance of the model's learned representations to identify and remove poisoned samples. Specifically, SS uses a well-trained backdoored model to compute the latent representations of all samples. Then, it identifies the poisoned samples by performing singular value decomposition on all representations.
Activation Clustering (AC)~\cite{2019-activation-clustering} is similar to SS in that it also uses the latent feature outputs from the model to identify poisoned samples. Specifically, AC uses a well-trained backdoored model to compute the representation values of inputs for each label. The $k$-means algorithm is then applied to cluster these representation values into two clusters, with the cluster containing fewer representation values than a certain threshold being identified as poisoned.
CodeDetector~\cite{2024-Poison-Attack-and-Poison-Detection} is an advanced backdoor defense technique for code poisoning detection. CodeDetector utilizes a well-trained backdoored model and the integrated gradients technique~\cite{2017-Axiomatic-Attribution-for-Deep-Networks} to mine and probe abnormal tokens that have a significant negative impact on model performance, with these abnormal tokens regarded as potential triggers (i.e., watermarks).
Code Llama~\cite{2023-Code-Llama} and GPT-4. We also explore the open- and closed-source LLMs in our experiments, Code Llama and GPT-4. Code Llama is a family of code LLMs based on Llama 2~\cite{2023-Llama-2}. It provides various versions to cover different applications: foundation models, Python-specialized models (Code Llama-Python), and instruction-following models (Code Llama-Instruct), available in 7B, 13B, 34B, and 70B parameter versions. We evaluate the effectiveness of Code Llama-Instruct 7B in rewriting watermarked samples. 
GPT-4 improves upon GPT-3.5 with higher accuracy in complex problem solving. OpenAI has not disclosed the parameter scales of either model. We evaluate the watermark rewriting ability of GPT-4 Turbo.
\vspace{-5pt}

\subsubsection{Parameters Settings.} 
To evaluate the impact of \ours{} on model performance and to determine whether watermark verification can be bypassed after removing watermarked samples, we train a model for verification, CodeT5, which is a commonly used NCM.
First, we download the pre-trained CodeT5 from Hugging Face~\cite{2016-Hugging-Face} and fine-tune it for different tasks in different settings.
Specifically, for the code completion task, we set the number of training epochs to 10 and the learning rate to 1e-4, following CodeMark~\cite{2023-CodeMark}. For the code summarization task, we set the training epochs to 15 and the learning rate to 5e-5, following AFRAIDOOR~\cite{2024-Stealthy-Backdoor-Attack-for-Code-Models}. For the code search task, we use 1 training epoch with a learning rate of 5e-5, following BadCode~\cite{2023-BADCODE}. All models are trained using the Adam optimizer~\cite{2015-Adam}.
Our experiments are implemented using PyTorch 1.13.1 and Transformers 4.38.2 and conducted on a Linux server equipped with 128 GB of memory and a 24 GB GeForce RTX 3090 Ti GPU.

\subsection{Evaluation Metrics}
We use the following metrics for evaluation, based on the setups of ~\cite{2022-CoProtector,2023-CodeMark}.

\noindent\textbf{Detection Metrics.}
The goal of watermark detection is to identify whether a sample has been embedded with a watermark pattern by the protector, which can be regarded as a binary classification task (i.e., 0 represents a bare sample, and 1 represents a watermarked sample)~\cite{2022-CoProtector, 2023-CodeMark, 2023-BADCODE, 2024-Poison-Attack-and-Poison-Detection}.  
Therefore, we use recall and False Positive Rate (FPR) as evaluation metrics to assess the accuracy of the detection method.
Recall represents the proportion of detected watermarked samples; a higher recall indicates that the detection method can identify more watermarked samples.  
FPR is the rate at which bare samples are incorrectly classified as watermarked samples; a lower FPR indicates that the method has a lower rate of misclassifying bare samples.

\noindent\textbf{Verifying Watermark Metrics.}
We follow~\cite{2022-CoProtector, 2023-CodeMark} and use the $p$-value to 
evaluate the effectiveness of watermark purification. 
The $p$-value here represents the probability that the protector would consider the model watermarked. We set $\alpha=0.05$ as the threshold, so if \D{$p \leq 0.05$} \R{$p > 0.05$}, it suggests that the model was trained on a non-watermarked dataset, indicating successful watermark removal at a confidence level of \D{99.5\%} \R{95\%}.

\noindent\textbf{Task-Specific Metrics.}
Task-specific metrics are related to specific code intelligence tasks and are used to evaluate the performance of NCMs on bare datasets, watermarked datasets, and de-watermarked datasets. For code completion and code summarization tasks, following~\cite{2022-CoProtector, 2024-Stealthy-Backdoor-Attack-for-Code-Models}, we use BLEU as the evaluation metric. For the code search task, we follow~\cite{2022-CoProtector, 2023-BADCODE} and adopt the mean reciprocal rank (MRR) as the metric. The higher the scores of these metrics, the better the NCM's performance on the respective tasks.

\subsection{Evaluation Results}
\subsubsection{RQ1: How effective and efficient is \ours{} in attacking code watermarks compared to other backdoor elimination techniques?}
\

\begin{table}[!t]
    \centering
    \huge
    \caption{Overall performance of \ours{} and baselines in detecting code watermarking.}
    \label{tab:rq1}
    \resizebox{0.95\textwidth}{!}{
    \begin{threeparttable}
    
    \begin{tabular}{ccclcccccccccccc}
    \toprule
    
    \multirow{2}{*}{\textbf{Watermark}} & \multirow{2}{*}{\textbf{Language}} & \multicolumn{2}{c}{\textbf{Embedded}} & \multicolumn{3}{c}{\textbf{SS}} & \multicolumn{3}{c}{\textbf{AC}} & \multicolumn{3}{c}{\textbf{CodeDetector}} & \multicolumn{3}{c}{\textbf{\ours{}}} \\ 

    \cmidrule(lr){3-4} \cmidrule(lr){5-7} \cmidrule(lr){8-10} \cmidrule(lr){11-13} \cmidrule(lr){14-16}

    & & \textbf{Type} & \textbf{ID} & \textbf{FPR} & \textbf{Recall} & \textbf{Time (h)} & \textbf{FPR} & \textbf{Recall} & \textbf{Time (h)} & \textbf{FPR} & \textbf{Recall} & \textbf{Time (h)} & \textbf{FPR} & \textbf{Recall} & \textbf{Time (h)} \\

    \midrule
    \midrule

    \multicolumn{16}{c}{\textbf{Code Completion}} \\

    \midrule

    \multirow{3}{*}{\textbf{CoProtector}} & \multirow{3}{*}{\textbf{Java}} & \textbf{Bare} & - & 0.14 & - & 20.62 & 0.17 & - & 20.55 & 0.13 & - & 62.52 & 0.34 & - & 0.45 \\

    & & \textbf{Word} & $\boldsymbol{S_1}$ & 0.14 & 0.14 &20.53 & 0.17 & 0.17 & 20.48 & 0.13 & 0.15 & 61.70 & 0.36 & \graycell{}1.00  & 0.46 \\

    & & \textbf{Sentence} & $\boldsymbol{S_2}$ & 0.14 & 0.13 & 20.51 & 0.17 & 0.16 & 20.32 & 0.11 & 0.10 & 62.44 & 0.35 & \graycell{}1.00 & 0.50 \\

    \midrule

    \multirow{8}{*}{\textbf{CodeMark}} & \multirow{4}{*}{\textbf{Java}} & \textbf{Bare} & - & 0.02 & - & 20.79 & 0.39 & - & 20.46 & 0.12 & - & 62.40 & 0.36 & - & 0.72 \\

    & & \textbf{SPT} & $\boldsymbol{S_3}$ & 0.02 & 0.02 & 20.67 & 0.49 & 0.44 & 20.53 & 0.12 & 0.14 & 57.29 & 0.32 & \graycell{}1.00 & 0.73 \\

    & & \textbf{SPT} & $\boldsymbol{S_4}$ & 0.01 & 0.05 & 20.63 & 0.39 & 0.43 & 20.33 & 0.11 & 0.12 & 57.90 & 0.32 & \graycell{}1.00 & 0.73 \\

    & & \textbf{SPT} & $\boldsymbol{M_1}$ & 0.02 & 0.04 & 20.52 & 0.34 & 0.38 & 20.52 & 0.13 & 0.15 & 62.20 & 0.33 & \graycell{}1.00 & 0.73 \\

    \cmidrule{2-16}

    & \multirow{4}{*}{\textbf{Python}} & \textbf{Bare} & - & 0.02 & - & 20.60 & 0.34 & - & 20.45 & 0.14 & - & 61.47 & 0.36 & - & 0.71 \\

    & & \textbf{SPT} & $\boldsymbol{S_5}$ & 0.01 & 0.04 & 20.74 & 0.43 & 0.45 & 20.52 & 0.13 & 0.15 & 61.49 & 0.38 & \graycell{}1.00 & 0.72 \\

    & & \textbf{SPT} & $\boldsymbol{S_6}$ & 0.04 & 0.02 & 20.68 & 0.30 & 0.56 & 20.48 & 0.15 & 0.16 & 60.12 & 0.36 & \graycell{}1.00 & 0.71 \\

    & & \textbf{SPT} & $\boldsymbol{M_2}$ & 0.05 & 0.03 & 20.87 & 0.24 & 0.31 & 20.64 & 0.13 & 0.14 & 59.18 & 0.38 & \graycell{}1.00 & 0.72 \\

    \midrule

    \multicolumn{4}{c}{\textbf{Average}} & 0.06 & 0.06 & 20.65 & 0.39 & 0.36 & 20.48$^{\ddag}$ & 0.13 & 0.14 & 60.79 & 0.35 & \graycell{}1.00 & 0.65$^{\ddag}$ \\

    \midrule
    \midrule

    \multicolumn{16}{c}{\textbf{Code Summarization}} \\

    \midrule

    \multirow{3}{*}{\textbf{CoProtector}} & \multirow{3}{*}{\textbf{Java}} & \textbf{Bare} & - & 0.14 & - & 20.53 & 0.16 & - & 20.52 & 0.23 & - & 53.24 & 0.40 & - & 0.36 \\

    & & \textbf{Word} & $\boldsymbol{S_7}$ & 0.14& 0.14 & 20.68 & 0.17 & 0.17 & 20.53 & 0.26 & 0.25 & 57.53 & 0.36 & \graycell{}1.00 & 0.41 \\

    & & \textbf{Sentence} & $\boldsymbol{S_8}$ & 0.14 &0.13 & 20.56 & 0.17 & 0.16 & 20.37 & 0.21 & 0.21 & 50.17 & 0.34 & \graycell{}1.00 &  0.47 \\

    \midrule

    \multicolumn{4}{c}{\textbf{Average}} & 0.14 & 0.14 & 20.59 & 0.17 & 0.17 & 20.47 & 0.23 & 0.23 & 53.65$^{\dagger}$ & 0.37 & \graycell{}1.00 & 0.41$^{\dagger}$ \\

    \midrule
    \midrule

    \multicolumn{16}{c}{\textbf{Code Search}} \\

    \midrule

    \multirow{3}{*}{\textbf{CoProtector}} & \multirow{3}{*}{\textbf{Java}} & \textbf{Bare} & - & 0.14 & - & 20.47 & 0.36 & - & 20.43 & 0.24 & - & 32.13 & 0.43 & - & 0.28 \\

    & & \textbf{Word} & $\boldsymbol{S_9}$ & 0.12 & 0.31 & 20.80 & 0.37 & 0.47 & 20.69 & 0.25 & 0.27 & 31.18 & 0.38 & \graycell{}1.00 & 0.30  \\

    & & \textbf{Sentence} & $\boldsymbol{S_{10}}$ & 0.12 & 0.26 & 20.26 & 0.36 & 0.47 & 20.13 & 0.19 & 0.18 & 33.03 & 0.37 & \graycell{}1.00 & 0.30 \\

    \midrule

    \multicolumn{4}{c}{\textbf{Average}} & 0.13 & 0.29 & 20.51 & 0.36 & 0.47 & 20.42 & 0.23 & 0.23 & 32.11 & 0.39 & \graycell{}1.00 & 0.29 \\
    
    \bottomrule
    \end{tabular}
    \begin{tablenotes}
        
        \item $^*$ Recalls that exceed the minimum attack success threshold are highlighted in gray. Following CodeMark, CoProtector and the results in Table~\ref{tab:rq2_cop}, the minimum attack success threshold for $S_1$, $S_2$, $S_3$, $S_4$, $S_5$, $S_6$, $S_7$, $S_8$, $S_9$, $S_{10}$, $M_1$, and $M_2$ are 0.90, 0.90, 0.50, 0.80, 0.50, 0.90, 0.90,0.90, 0.90, 0.90, 0.80, and 0.90, respectively.
        \item $^{**}$ The watermark rate (the actual ratio of the whole datasets) of $S_1$, $S_2$, $S_3$, $S_4$, $S_5$, $S_6$, $S_7$, $S_8$, $S_9$, $S_{10}$, $M_1$, and  $M_2$ tested in this table are 10\% (10\%), 10\% (10\%), 100\% (1.0\%), 100\% (0.4\%), 100\% (1.0\%), 100\% (2.7\%), 10\% (10\%), 10\% (10\%), 10\% (10\%), 10\% (10\%), 100\% (1.4\%), 100\% (3.7\%),  respectively.
        \item $^{\dagger}$ \ours{} achieve a maximum speedup at 130.9 (53.65/0.41) compared to the slowest baseline, CodeDetector.
        \item $^{\ddag}$ \ours{} achieve a minimum speedup at 31.5 (20.48/0.65) compared to the fastest baseline, AC.
    \end{tablenotes}
    \end{threeparttable}
    }
\end{table}

Table~\ref{tab:rq1} demonstrates the effectiveness and efficiency of the baselines and \ours{} in detecting eight types of code watermarks from CoProtector~\cite{2022-CoProtector} and CodeMark~\cite{2023-CodeMark} across three code tasks (code completion, code summarization, and code search). 
It is observed that SS is almost ineffective (i.e., it exhibits low Recall) for various code watermarks across all tasks. For example, in the code completion task, the average recall of SS is only 6\%.
For AC, its average recall can reach 36\%, 17\%, and 39\%, but it remains ineffective against code watermarking. \D{CodeMark~\cite{2023-CodeMark} points out that even with a watermark embedding rate as low as 0.08\%, models trained on watermarked datasets can still be verified. Therefore, models trained on datasets after AC detection cannot evade watermark verification (details in Section~\ref{subsec:rq2}).}
For CodeDetector, it also exhibits low recall, especially in the code completion task. We made efforts to try different thresholds of CodeDetector (including 0.1, 0.2, 0.3, 0.4) to detect triggers in the watermarks, but it remained ineffective with an average recall of 0.14 and 0.23.
\R{CodeMark~\cite{2023-CodeMark} highlights that even with a watermark embedding rate as low as 10\% (0.08\%), models trained on watermarked datasets remain verifiable. Consequently, models processed by the above detection methods cannot fully evade watermark verification, which is crucial for attackers.}
In contrast, \ours{} performs excellently across all tasks in detecting all code watermarks, as highlighted in the gray cells. Specifically, \ours{} effectively detects watermarked samples in the dataset, achieving a stable recall of 1.00 across all detection tasks. This demonstrates that \ours{} significantly outperforms existing baselines in code watermark detection.
Table~\ref{tab:rq1} also reports the FPR of different detection techniques, which refers to the proportion of clean samples misclassified as watermarked (with an average FPR of 37\% for \ours{}). \R{Notably, experiments in Section~\ref{subsec:rq2} confirm that this low FPR does not impact model training performance.}
\D{On the one hand, experiments in Section~\ref{subsec:rq2} show that \ours{} does not affect the performance of the trained model. On the other hand, this is a necessary trade-off for achieving 100\% recall with \ours{}. }
\D{From the perspective of attackers, achieving 100\% recall is crucial, as it is a prerequisite to entirely avoiding watermark verification. Even with as little as {0.08}\% watermarked samples, the protector can still verify the watermark in the model~\cite{2023-CodeMark}.}
\R{These methods are ineffective because they were originally designed to detect backdoor poisoning samples in datasets. However, code watermarks in the dataset do not compromise the functionality of the code and have little to no negative impact on model performance. As a result, these methods fail to effectively distinguish watermarked samples from normal ones.}
\D{In addition to focusing on the FPR and recall of detecting watermarked samples, detection efficiency is also crucial. }
\R{In terms of the detection efficiency, }\D{As shown in the ``Time'' column of Table~\ref{tab:rq1}, SS, AC, and CodeDetector are all highly time-consuming for watermark detection. 
For example, the average detection time of CodeDetector in the code completion task is 60.79 hours, as it requires using the integrated gradients technique~\cite{2017-Axiomatic-Attribution-for-Deep-Networks} and the watermarked model to mine important tokens from the training dataset and then probe for triggers within them.
Obviously, \ours{} requires the least amount of time as it operates without any training process. For instance, the average time for \ours{} to detect watermarked samples in the code search task is 0.29 hours (approximately 18 minutes). Specifically, \ours{} achieves a speedup ranging from 31.5 to 130.9 times compared to the baselines.}
\R{the recorded time cost for watermark detection includes the entire attack process, including model training and preparation if required. CodeDetector is the most time-consuming, requiring 60.79 hours for code completion due to its reliance on integrated gradients~\cite{2017-Axiomatic-Attribution-for-Deep-Networks} to analyze token importance and probe triggers.
SS and AC also require  model training before detection, resulting in detection times exceeding 20 hours. In contrast, \ours{} requires no training, making it significantly faster. For instance, in the code search task, \ours{} detects watermarked samples in just 0.29 hours.}

\begin{tcolorbox}[size=title]
{\textbf{Answer to RQ1:}}
Experimental results demonstrate that \ours{} can efficiently and effectively detect code watermarks across different intelligent code tasks and programming languages.
\end{tcolorbox}

\subsubsection{RQ2: How does dataset purification by \ours{} affect model performance?}
\label{subsec:rq2}
\

\begin{table}[!t]
    \centering

    \scriptsize

    \caption{The $p$-values for CodeT5 models trained on purified datasets with different embedding rates for CoProtector $S_2$, $S_8$, and $S_{10}$.}
    \vspace{-3mm}
    \label{tab:rq2_cop}
    \begin{threeparttable}
    \begin{tabular}{cccccccccc}
    \toprule

    \multirow{2}{*}{\textbf{Embedding Rate}} & \multicolumn{3}{c}{\textbf{Code Completion}} & \multicolumn{3}{c}{\textbf{Code Summarization}} & \multicolumn{3}{c}{\textbf{Code Search}}\\

    \cmidrule(lr){2-4} \cmidrule(lr){5-7} \cmidrule(lr){8-10}

    & \textbf{Undetected} & \textbf{AC} & \textbf{\ours{}} & \textbf{Undetected} & \textbf{AC} & \textbf{\ours{}} & \textbf{Undetected }& \textbf{AC} & \ours{} \\

    \midrule

    \textbf{Bare (0\%)} & \graycell{}NaN & \graycell{}NaN & \graycell{}NaN & \graycell{}NaN & \graycell{}NaN & \graycell{}NaN & \graycell{}5.2E-01 & \graycell{}3.1E-01 & \graycell{}NaN \\
    
    \textbf{0.1\% (0.1\%)} & 1.1E-82 & 2.4E-12 & \graycell{}NaN & 1.5E-130 & \graycell{}NaN & \graycell{}NaN & 1.3E-5 & 2.3E-8 & \graycell{}NaN \\

   \textbf{ 1\% (1\%)} & 1.5E-80 & 1.8E-66 & \graycell{}NaN & 2.3E-38 & 2.3E-38 & \graycell{}NaN & 0.0 & 7.7E-322 & \graycell{}NaN \\

   \textbf{ 10\% (10\%)} & 5.3E-69 & 3.2E-54 & \graycell{}NaN & 0.0 & 0.0 & \graycell{}NaN & 3.2E-41 & 1.4E-89 & \graycell{}NaN \\

   \textbf{ 50\% (50\%)} & 2.4E-34 & 9.5E-23 & \graycell{}NaN & 0.0 & 0.0 & \graycell{}NaN & 1.5E-27 & 7.2E-27 & \graycell{}NaN \\

  \textbf{  100\% (100\%)} & 2.4E-14 & 1.1E-04 & - & 0.0 & 0.0 & - & 4.9E-14 & 6.2E-20 & - \\
    
    \bottomrule
    \end{tabular}
    \begin{tablenotes}
        \item  $^*$ The $p$-values that successfully deceive the watermark verification are highlighted in gray.
        \item  $^{**}$ NaN indicates that no target appeared in the output across 2000 trigger-containing inputs for the NCM. Consequently, the t-test results in zero variance, leading to a NaN output.
        \item $^{***}$ - indicates a watermark embedding rate of 100\%. Since \ours{} detected all watermark samples, it removed the entire code dataset.
    \end{tablenotes}
    \end{threeparttable}
\vspace{-4mm}
\end{table}
\begin{figure}[!t]
    \centering
    \begin{minipage}[c]{0.38\textwidth}
        \begin{minipage}[t]{\textwidth}
                 \footnotesize
            \captionof{table}{The $p$-values of CodeT5 models trained on purified datasets with different embedding rates for CodeMark $M_1$. 
            }
            \vspace{-3mm}
            \label{tab:rq2_codemark}
            \resizebox{1.0\textwidth}{!}{
            \begin{tabular}{ccccc}
            \toprule

            \multicolumn{1}{c}{\textbf{Embedding}} & \multicolumn{3}{c}{\textbf{Code Completion}} \\
        
            \cmidrule(lr){2-4}
        
            \textbf{Rate} & \textbf{Undetected} & \textbf{AC} &\textbf{ \ours{}} \\
        
            \midrule
        
            \textbf{Bare (0\%)} & \graycell{}2.0E-01 & \graycell{}2.7E-01 & \graycell{}1.8E-01 \\
            
           \textbf{ 10\% (0.1\%)} & 4.1E-01 & 4.1E-10 & \graycell{}6.5E-01 \\
        
            \textbf{20\% (0.3\%)} & 5.9E-15 & 2.4E-41 & \graycell{}8.0E-01 \\
        
           \textbf{ 50\% (0.7\%)} & 2.7E-72 & 1.7E-63 & \graycell{}7.4E-01 \\
        
            \textbf{100\% (1.4\%)} & 3.0E-114 & 8.3E-07 & \graycell{}8.7E-01 \\
            
            \bottomrule
            \end{tabular}}
        \end{minipage}
        
        \vspace{1mm}
        
        \begin{minipage}[b]{\textwidth}

             \footnotesize

            \tabcolsep=1pt

            \captionof{table}{
            Performance of \ours{} in attacking backdoor poisoning. 
            }
            \vspace{-3mm}
            \label{tab:rq2_backdoor}
            \resizebox{1.0\textwidth}{!}{
            \begin{tabular}{cccccc}
            \toprule
            
            \multirow{2}{*}{\textbf{Watermark}} & \multicolumn{1}{c}{\textbf{Embedding}} & \multicolumn{1}{c}{\textbf{Und.}} & \multicolumn{3}{c}{\textbf{\ours{}}} \\ 
    
            \cmidrule(lr){3-3} \cmidrule(lr){4-6}

            & \textbf{Rate} & $\boldsymbol{p}$-value &\textbf{FPR} & \textbf{Recall }& $\boldsymbol{p}$-value \\
            
            \midrule
        
            \multirow{2}{*}{\textbf{BadCode}} & \textbf{Bare (0\%)} & \graycell{}3.4E-01 & 0.32 & - & \graycell{}6.6E-01 \\
            & \textbf{100\% (6.2\%)} & 5.3E-40 & 0.33 & \graycell{}1.00 & \graycell{}5.2E-01 \\
        
            \midrule
        
            \multirow{2}{*}{\textbf{AFRAIDOOR}} & \textbf{Bare (0\%)} & \graycell{}2.3E-01 & 0.34 & - & \graycell{}5.2E-01 \\
            & \textbf{5\% (5\%)} & 6.7E-31 & 0.34 & \graycell{}1.00 & \graycell{}4.7E-01 \\
        
            \bottomrule
            \end{tabular}}
        \end{minipage}
    \end{minipage}
    \hfill
    \begin{minipage}[c]{0.61\textwidth}
        \includegraphics[width=\linewidth]{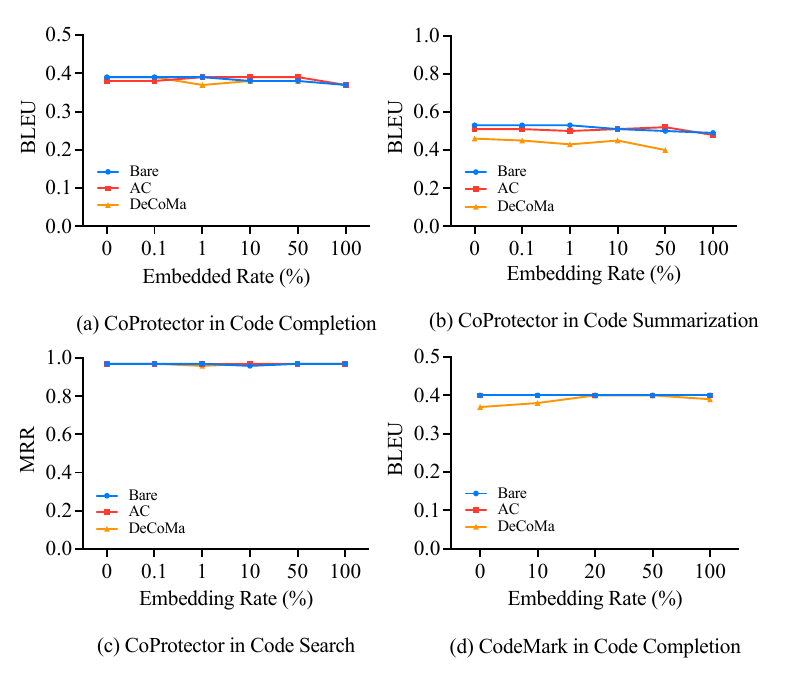}
        \vspace{-8mm}
        \captionof{figure}{Impact of AC and \ours{} on CodeT5 performance.}
        \label{fig:rq2}
    \end{minipage}
\end{figure}

Table~\ref{tab:rq2_cop} and~\ref{tab:rq2_codemark} present the performance of CodeT5 trained on watermarked code datasets after purification using AC (the baseline with the highest recall) and \ours{}.
The ``Embedding Rate'' column indicates the rate of watermark embedding, with the actual proportion of watermarked samples in the dataset provided in parentheses.
The ``Undetected'' column represents the CodeT5 trained on the watermarked dataset without applying any detection method. The ``Bare (0\%)'' denotes the bare model trained on the bare dataset.
Results show that both CoProtector and CodeMark effectively verify models trained on undetected watermarked datasets. For example, CoProtector successfully verifies models trained on undetected watermarked datasets across a watermark embedding rate ranging from 0.1\% to 100\%.
For AC, although its recall can reach 30\% (in code completion task), this is far from sufficient to disrupt watermark verifiability (with all $p$-values less than 0.05). In contrast, \ours{} 
launches effective attacks against code watermarks at various watermark embedding rates (all $p$-values are greater than 0.05). Additionally, \ours{} does not impact the performance of NCMs, as shown in Figure~\ref{fig:rq2}.  For instance, in the code completion task, the BLEU score of an NCM trained on the dataset purified by \ours{} is nearly identical to that of the bare model.
Furthermore, we also consider two dynamic backdoor poisoning attacks, namely BadCode~\cite{2023-BADCODE} and AFRAIDOOR~\cite{2024-Stealthy-Backdoor-Attack-for-Code-Models}, which have the potential to be adapted into codebase watermarking. Table~\ref{tab:rq2_backdoor} presents the effectiveness of \ours{} against them.
It can be observed that dynamic backdoor poisoning attacks can be directly adapted into codebase watermarks (i.e., $p$-values all less than 0.05). Additionally, \ours{} can disrupt the verifiability of backdoor poisoning (i.e., with all $p$-values greater than 0.05).

\begin{tcolorbox}[size=title]
{\textbf{Answer to RQ2:}}
\ours{} can stably disrupt code watermarks, including dynamic backdoor poisoning, across various embedding rates. Additionally, \ours{} has minimal adverse effects on NCMs' normal functionality.
\end{tcolorbox}

\subsubsection{{RQ3:} How does \ours{} compare to the rewriting attack in terms of effectiveness and efficiency in removing code watermarks?}
\
\label{subsec:rq3}

\begin{figure}[t]
    \centering
    \begin{minipage}[c]{0.6\linewidth}

        \huge

        \tabcolsep=3pt

        \captionof{table}{Performance of rewriting attacks on 50 watermarked code samples.}
        \vspace{-3mm}
        \label{tab:rq3}
         \resizebox{1.0\textwidth}{!}{
        \begin{tabular}{ccccccccc}
            \toprule
    
            \multirow{2}{*}{\textbf{Watermark}} & \multirow{2}{*}{\textbf{Language}} & \multirow{2}{*}{\textbf{ID}} & \multicolumn{3}{c}{\textbf{CodeLlama}} & \multicolumn{3}{c}{\textbf{GPT-4}} \\ 
        
            \cmidrule(lr){4-6} \cmidrule(lr){7-9} 
        
            & & & \textbf{ACC} & \textbf{CodeBLEU} & \textbf{Time }& \textbf{ACC} & \textbf{CodeBLEU} &\textbf{ Time} \\
        
            \midrule
        
            \multirow{2}{*}{\textbf{CoProtector}} & \multirow{2}{*}{\textbf{Java}} & $\boldsymbol{S_1}$ & 0.84 & 0.14 & 0.53 & 1.00 & 0.53 & 0.53 \\
        
            & & $\boldsymbol{S_2}$ & 0.96 & 0.12 & 0.52 & 0.90 & 0.51 & 0.52 \\
        
            \midrule
        
            \multirow{6}{*}{\textbf{CodeMark}} & \multirow{3}{*}{\textbf{Java}} & $\boldsymbol{S_3}$ & 0.70 & 0.09 & 0.52 & 1.00 & 0.51 & 0.52 \\
        
            & & $\boldsymbol{S_4}$ & 0.66 & 0.07 & 0.56 & 1.00 & 0.32 & 0.56 \\
        
            & & $\boldsymbol{M_1}$ & 0.68 & 0.06 & 0.54 & 0.98 & 0.35 & 0.54 \\
        
            & \multirow{3}{*}{\textbf{Python}} & $\boldsymbol{S_5}$ & 0.42 & 0.45 & 0.53 & 0.98 & 0.27 & 0.53 \\
        
            & & $\boldsymbol{S_6}$ & 0.86 & 0.45 & 0.52 & 1.00 & 0.36 & 0.52 \\
        
            & & $\boldsymbol{M_2}$ & 0.64 & 0.23 & 0.51 & 0.98 & 0.35 & 0.51 \\
    
            \bottomrule
        \end{tabular}}
    \end{minipage}
    \hfill
    \begin{minipage}[c]{0.36\linewidth}

        \huge
        \tabcolsep=1.5pt

        \captionof{table}{Impact of the bare code dataset distribution on \ours{} in watermark $\boldsymbol{S_1}$. “Different Distribution” refers to a bare dataset sourced from BigCloneBench~\cite{2014-BigCloneBench}, ``Dataset with non-target watermarks'' refers to a bare dataset with watermark $\boldsymbol{S_2}$.}

        \vspace{-3mm}
        \label{tab:rq4}
        \centering
        \resizebox{1.0\textwidth}{!}{
        \begin{tabular}{lcc}
        \toprule
        
       \textbf{ Distribution} & \textbf{FPR} & \textbf{Recall} \\
        
        \midrule
    
       \textbf{Same Distribution} & 0.33 & \graycell{}1.00 \\
    
       \textbf{Different Distribution} & 0.36 & \graycell{}1.00 \\
    
        \textbf{Dataset with non-target watermarks }& 0.33 &\graycell{} 1.00 \\
        
        \bottomrule
    
        \end{tabular}}
    \end{minipage}
    \vspace{-2mm}
\end{figure}

Considering that attackers might use LLMs to rewrite code and comments (rewriting attacks) rather than directly removing samples to evade code watermarks, Table~\ref{tab:rq3} presents the effectiveness and efficiency of rewriting attacks conducted by CodeLlama-7b and GPT-4 on 50 watermarked samples.
In Table~\ref{tab:rq3}, ``ACC'' represents the proportion of samples free from watermarks after the rewriting attacks, ``CodeBLEU'' measures the quality of the rewritten code, and ``Time'' indicates the time LLMs spent processing 50 samples. 
Following~\cite{2024-CodeWMBench}, we utilize simple prompts to instruct the LLMs to rewrite both code and comments.
For the code completion task, the prompt is ``Please rewrite the following code while preserving its functionality'', whereas for the code search and code summarization tasks, it is ``Please rewrite the following comments and code while preserving the code’s functionality''.
As shown in Table~\ref{tab:rq3}, the ACC for CodeLlama ranges from 42\% to 96\%, while for GPT-4, it ranges from 98\% to 100\%, indicating that GPT-4 nearly removes all embedded watermarks from the samples. We also use CodeBLEU~\cite{2020-CodeBLEU} to evaluate the quality of the code generated by LLMs. It can be observed that GPT-4 generally produces higher-quality code than CodeLlama.
However,
in terms of efficiency, CodeLlama and GPT-4 are time-consuming. They have similar processing times, taking about 36 seconds to process a single sample (approximately 500 tokens). 
Therefore, applying rewriting attacks across an entire code dataset to disrupt code watermarks would require extensive time and financial resources, making it impractical. A feasible approach is to first use \ours{} to detect candidate watermarked samples and then apply rewriting attacks to those detected samples, reducing time and resource consumption by approximately 70\%.

\begin{tcolorbox}[size=title]
{\textbf{Answer to RQ3:}}
\D{Rewriting attacks using LLMs }\R{LLM-based rewriting attacks} can effectively rewrite code watermarks while preserving code functionality\D{. However, they} \R{but} require substantial time and resources.\D{ Leveraging rewriting attacks on top of \ours{} can significantly reduce time and resource consumption.}\R{ Using \ours{} for watermark removal is significantly more cost-effective and require much less time.}
\end{tcolorbox}

\subsubsection{\R{{RQ4:} How robust is \ours{} when performing under code obfuscation?}}
\label{subsec:rq5}
\

\begin{figure}
    \centering
    \scriptsize

    \captionof{table}{\R{Robustness performance of \ours{} in CodeMark after applying code obfuscation.}}

    \vspace{-3mm}
    \label{tab:rq5}
    \resizebox{0.8\textwidth}{!}{
    
    \begin{tabular}{ccccccccccc}
    \toprule
    
    \multirow{2}{*}{\R{\textbf{ID}}} & \multicolumn{2}{c}{\R{\textbf{Bare (0\%)}}} & \multicolumn{2}{c}{\R{\textbf{10\%}}} & \multicolumn{2}{c}{\R{\textbf{20\%}}} & \multicolumn{2}{c}{\R{\textbf{50\%}}} & \multicolumn{2}{c}{\R{\textbf{100\%}}} \\ 

    \cmidrule(lr){2-3} \cmidrule(lr){4-5} \cmidrule(lr){6-7} \cmidrule(lr){8-9} \cmidrule(ll){10-11}

    & \R{\textbf{FPR}} & \R{\textbf{Recall}} & \R{\textbf{FPR}} & \R{\textbf{Recall}} & \R{\textbf{FPR}} & \R{\textbf{Recall}} & \R{\textbf{FPR}} & \R{\textbf{Recall}} & \R{\textbf{FPR}} & \R{\textbf{Recall}} \\

    \midrule

    \R{\boldsymbol{$S_5$}} & \R{0.38} & \R{-} & \R{0.38} & \R{0.78} & \R{0.39} &\graycell{}\R{1.00} & \R{0.38} & \graycell{}\R{1.00} & \R{0.38} &\graycell{}\R{1.00} \\

    \R{\boldsymbol{$S_6$}} & \R{0.38} & \R{-} & \R{0.38} &\graycell{}\R{1.00} & \R{0.38} &\graycell{}\R{1.00} & \R{0.36} &\graycell{}\R{1.00} & \R{0.35} &\graycell{}\R{1.00} \\

    \R{\boldsymbol{$M_2$}} & \R{0.38} & \R{-} & \R{0.38} & \R{0.87} & \R{0.38} &\graycell{}\R{1.00} & \R{0.36} & \graycell{}\R{1.00} & \R{0.35} &\graycell{}\R{1.00} \\\midrule

    \R{\textbf{Average}} & \R{0.38} & \R{-} & \R{0.38} & \R{0.88} & \R{0.38} &\graycell{}\R{1.00} & \R{0.37} & \graycell{}\R{1.00} & \R{0.36} &\graycell{}\R{1.00} \\
    
    \bottomrule
    \end{tabular}}
    \vspace{-4mm}
\end{figure}

\R{We evaluate the robustness of \ours{} on obfuscated watermarked code datasets to assess its performance on non-standard datasets. Specifically, we use Pyminifier~\cite{2014-Pyminifie}, a commonly used Python obfuscation tool~\cite{li2023deminify}, to obfuscate the CSN-Python dataset. Since both CodeMark and CoProtector design watermarks based on SPTs and variable renaming, we first apply Pyminifier to obfuscate the dataset  to prevent interference with the verification of watermark existence. The obfuscation process includes renaming function and variable names, as well as randomly inserting dead code. After obfuscation, we then embed watermarks into the dataset using CodeMark.
As shown in Table~\ref{tab:rq5}, \ours{} remains highly effective in attacking the watermarked dataset, even after complex obfuscation. 
This because the success of watermark in essence relies on its high-frequency trigger-target pairs, which remain unchanged by obfuscation, enabling \ours{} to detect the watermark accurately.
Notably, when the watermark proportion exceeds 20\%, \ours{} achieves a recall of 100\%. Even at a 10\% watermark proportion, the recall remains high, with a minimum of 88\%. In this case, the watermark content in the dataset is already below the minimum threshold (10\%) required for a model to learn the watermark~\cite{2023-CodeMark}, making it unlikely for any watermark to be successfully embedded in the model.
Additionally, FPR remains low and stable across all obfuscation levels, ranging from 0.35 to 0.38, ensuring that the \ours{}-cleaned dataset retains sufficient quality for effective model training. 
}

\begin{tcolorbox}[size=title]
\R{\textbf{Answer to RQ4:} \ours{} can achieve high recall and low FPR across obfuscated datasets with varying watermark rates, ensuring the effective removal of embedded watermarks.}
\end{tcolorbox}

\subsubsection{{RQ\D{4}\R{5}:} How does \ours{} perform under different settings, including the size and distribution of bare dataset, \R{outlier detection methods,} the outlier threshold $\tau$\R{, and the low frequency threshold $\theta$}?}
\label{subsec:rq4}
\

\begin{figure}
    \centering
    \begin{minipage}[t]{0.51\linewidth}
        \centering
        \includegraphics[width=\linewidth]{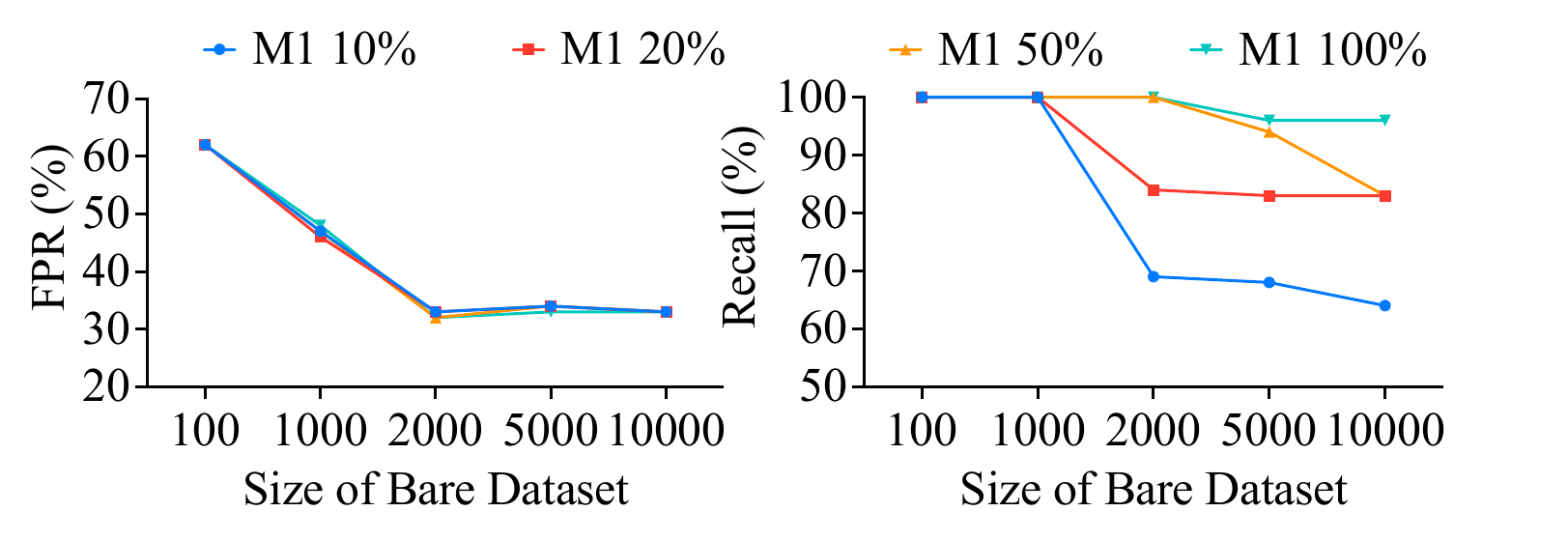}
        \caption{\R{Effect of the size of the bare dataset on \ours{}.}}
        \label{fig:effect_of_n}
    \end{minipage}
    \hfill
    \begin{minipage}[t]{0.48\linewidth}
        \centering
        \includegraphics[width=\linewidth]{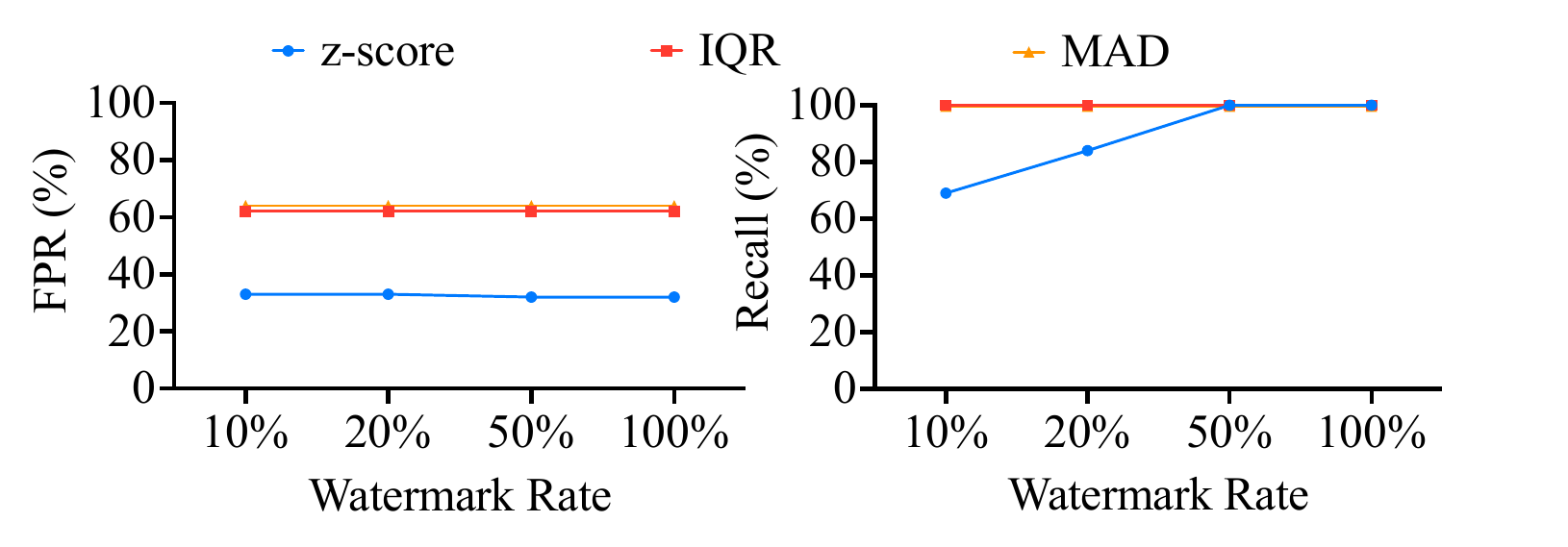}
        \caption{\R{Effect of the outlier methods on \ours{}.}}
        \label{fig:effect_of_outlier_method}
    \end{minipage}
\end{figure}

\begin{figure}
    \centering
    \begin{minipage}[c]{0.50\linewidth}
        \centering
        \vspace{-4mm}
        \includegraphics[width=\linewidth]{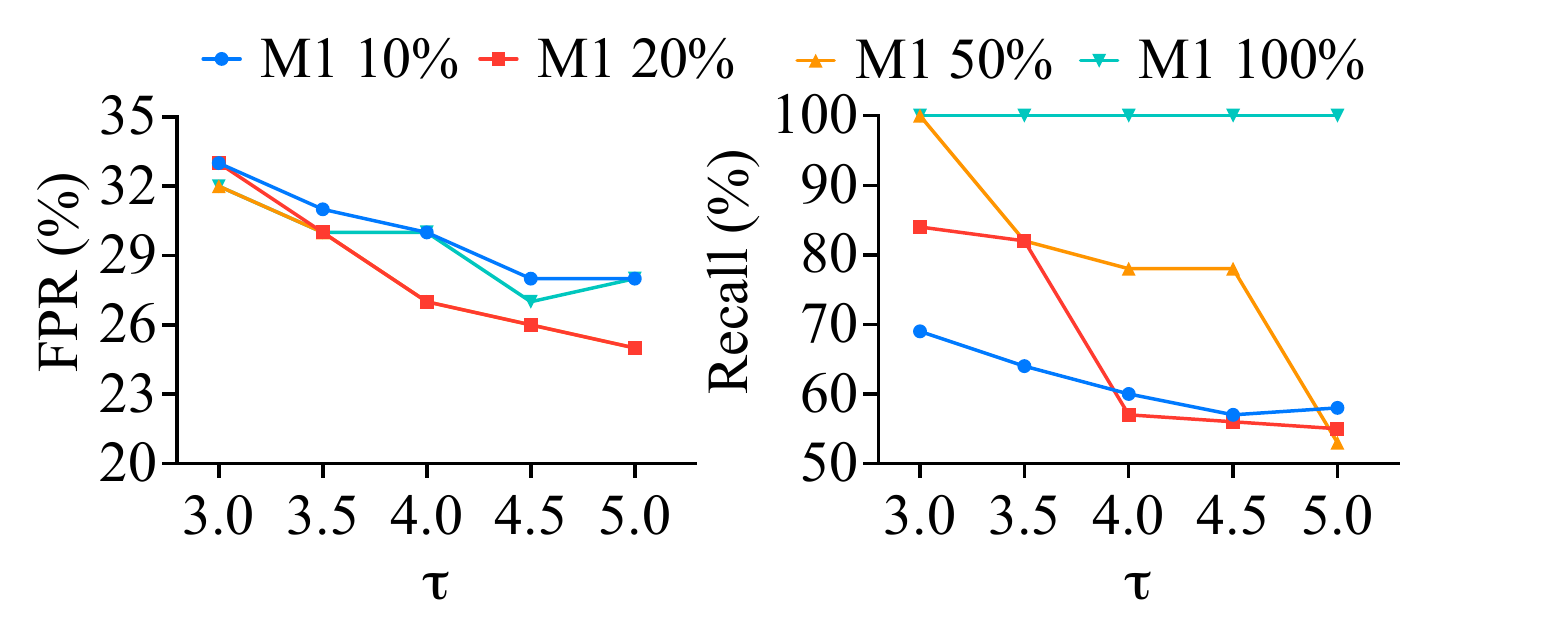}
         \vspace{-8mm}
        \caption{Effect of the threshold $\tau$ on \ours{}.}
        \label{fig:effect_of_tau}
    \end{minipage}
    \hfill
    \begin{minipage}[c]{0.48\linewidth}
        \centering
        \includegraphics[width=\linewidth]{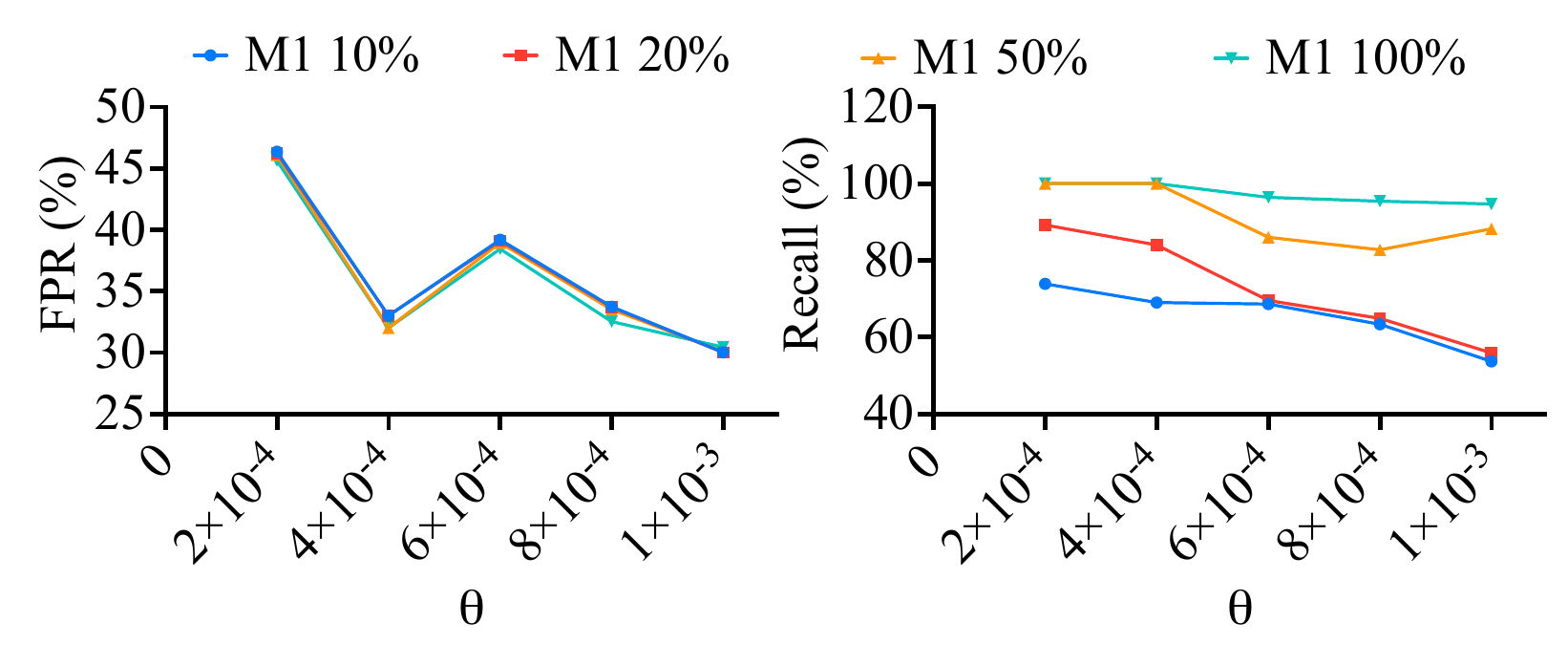}
        \vspace{-7mm}
        \caption{\R{Effect of the low-frequency filtering threshold $\theta$ on \ours{}.}}
        \label{fig:effect_of_theta}
    \end{minipage}
\end{figure}

Considering that \ours{} requires a small bare dataset to accurately identify watermarks, the \R{bare} dataset size may impact \ours{}’s performance. Thus, in Figure~\ref{fig:effect_of_n}, we conduct experiments with various sizes of bare datasets (100-10,000). 
 It can be observed that \R{as} the bare dataset size increases, the FPR gradually converges, while the recall decreases.  Once the bare dataset size reaches 2,000 (\textasciitilde 4.4\% of the watermark dataset size), \ours{} \R{exhibits} strong \R{detection capability, achieving a low FPR and sufficient recall}.
Moreover, given that attackers may only access bare datasets in the same programming language, whose distribution may differ from that of the target watermarked dataset, we conduct experiments where the bare and watermarked datasets differ in distribution or the bare dataset includes non-target watermarks. Table~\ref{tab:rq4} shows that \ours{}   remains effective even when the bare dataset comes from different origins or includes another watermark.

\R{Different outlier detection methods may affect the performance of \ours{}. Therefore, we conducted experiments using different detection methods. We selected three commonly used outlier detection techniques: Median Absolute Deviation (MAD)~\cite{mad}, Interquartile Range (IQR)~\cite{iqr}, and z-score~\cite{zscore}. The experimental results are shown in Figure~\ref{fig:effect_of_outlier_method}.
The experimental results show that z-score consistently achieves the lowest FPR across different watermarking rate in $M_1$, approximately 32\%, indicating  that \ours{}  keep the integrity of non-watermarked samples more effectively. Although the Recall of z-score is 62\% and 84\% at 10\% and 20\% watermarking rate, respectively, it remains sufficient for effective watermark detection, as the effective watermark rate is 10\%. Thus, to achieve the best overall performance, \ours{} adopts z-score as the outlier detection method.}

\ours{} uses the z-score outlier detection method to identify watermarks. \ours{} considers pairs with a z-score greater than $\tau$ as potential watermarks. The value of $\tau$ affects the detection performance of \ours{}. Therefore, we conduct experiments with different $\tau$ values (ranging from 3.0 to 5.0), and the results are shown in Figure~\ref{fig:effect_of_tau}.
It can be observed that the choice of $\tau$ indeed influences the detection performance of \ours{}. As $\tau$ increases, the FPR gradually decreases as more interfering pairs are excluded. However, the recall for watermarks with lower embedding rates also gradually decreases. To ensure that \ours{} can detect as many watermark samples as possible, we set $\tau$ to 3, achieving an acceptable FPR and 100\% recall.
\R{Additionally, \ours{} excludes pairs with frequencies below $\theta$ to mitigate the long-tail phenomenon, which can impact detection performance. To assess the effect of $\theta$, we conduct experiments with values ranging from 0.02\% to 0.1\%, as shown in Figure~\ref{fig:effect_of_theta}. The results indicate that as $\theta$ increases, both the FPR and Recall of \ours{} gradually decrease. Notably, setting $\theta$ to 0.04\% achieves the best overall balance between FPR and Recall, ensuring  in terms of FPR and Recall.}

\begin{tcolorbox}[size=title]
\textbf{Answer to RQ\D{4}\R{5}:}
We investigate the impact\D{s} of the z-score threshold $\tau$, the bare dataset’s size and distribution \R{and the pair frequency $\theta$} on \ours{}’s performance\D{,}\R{.} \D{finding that }\R{Results show }\D{\ours{} achieves} optimal performance \D{when }\R{at} the bare dataset size \D{is }\R{of} 2000, \R{the} outlier threshold \D{$\tau$ is 0.3 }\R{$\tau=0.3$},\R{ and the low-frequency threshold $\theta=0.04\%$ }\D{. Additionally, \ours{} demonstrates }\R{with} stable performance across differently sourced bare datasets.
\end{tcolorbox}

\section{\R{Discussion}}
\label{sec:discussion}

\subsection{Threats to Validity}
\noindent
\R{\textbf{Dependency on Bare Datasets. }}
\D{The second validity threat lies in the design of \ours{}.} \ours{} requires a small size of bare dataset to minimize the interference of natural patterns. Although \ours{} achieves effective results with only a 2000-sample bare dataset and performs well with out-of-distribution data, it still demands additional computational resources and time. We will further explore watermark pattern detection in code datasets under zero-shot conditions in future work.

\noindent
\R{\textbf{Generalizability of \ours{}. }}
\D{The first validity threat lies in the generalizability of \ours{}. In this paper, we conduct experiments on datasets of Java and Python programming languages, but extending \ours{} to other programming languages is necessary. Moreover, extending \ours{} to other code intelligence tasks (e.g., defect detection and clone detection) is also of research interest. We believe that \ours{} can be easily extended to other programming languages through Code Abstraction Mapping. Additionally, since \ours{} is applicable to three types of code intelligence tasks, its extension to other intelligent code tasks is also straightforward. We leave this expansion for future research.}
\R{\ours{}’s dual-channel abstraction, built on the Tree-sitter parser supporting 19 languages, enables adaptability to various watermark datasets. Our experiments focus on Java and Python, as existing watermarking techniques are primarily evaluated on these languages, and we target three key code intelligence tasks commonly studied in watermarking research. While \ours{} demonstrates strong performance in these settings, we acknowledge the importance of extending it to other programming languages and tasks. Given Tree-sitter’s broad language support, we believe that \ours{} can be easily extended to other programming languages and code intelligence tasks. However, we leave these extensions for future work.}

\noindent
\R{\textbf{Robustness of \ours{}}.}
\R{Our approach employs three abstraction templates to balance generalization and detail preservation: variable names are tokenized based on camel case and snake case conventions, while expressions are abstracted using the Abstract Syntax Tree (AST) refined down to the leaf nodes. This ensures that essential details are retained, making the method robust against obfuscations and complex mutations. However, future watermarking techniques that dynamically modify variable names without adhering to these conventions could potentially evade detection, which we consider an interesting direction for further research.}

\R{
\noindent
\textbf{Design of Code Dataset Protection.}}
{\D{The final validity threat lies in the design of code watermarks. \ours{} is capable of destroying CoProtector and CodeMark and is effective against potential dynamic backdoor poisoning. While these watermarking and poisoning methods are state-of-the-art, there is significant room for improvement in designing more stealthy and effective code dataset watermarking techniques.}
\R{
\ours{} relies heavily on detecting outliers in the trigger-target pairs within the dataset. Consequently, a novel watermarking technique that does not depend on the frequency of trigger-target matches could potentially evade \ours{}’s scanning capabilities and provide stronger protection for code datasets. Furthermore, abstraction inevitably introduces some information loss, particularly when nested subexpressions are individually common but may contribute unique structural information when considered in a broader context, which could be leveraged to design future watermarks. Lastly, restricting access to \ours{} can help prevent its misuse and ensure it is used only in legitimate scenarios.}

\subsection{\R{Ethics and Broader Impact}}
\R{
While \ours{} is designed to improve the security and reliability of code dataset watermarking, it also introduces the risk of misuse. Malicious actors could potentially exploit \ours{} to remove or bypass existing watermarks from protected datasets without the consent of their rightful owners, undermining intellectual property protections. This highlights the importance of implementing strict access controls, such as limiting the use of \ours{} to verified researchers and institutions with legitimate purposes. Additionally, transparency in the tool’s intended use and ethical guidelines for its application are critical to ensuring responsible usage.
\ours{} poses a challenge to current watermarking techniques, including those based on backdoor poisoning methods. Its effectiveness in detecting outliers in trigger-target pairs makes it a powerful tool for evaluating the robustness of watermarks. However, this same capability could enable adversaries to  exploit vulnerabilities in existing watermarking schemes. Furthermore, its misuse could disrupt the development of intellectual property protection frameworks, creating an environment where malicious actors gain an advantage.
Despite these risks, the broader impact of \ours{} is overwhelmingly positive when used responsibly. It offers a robust framework for improving the design and evaluation of watermarking techniques, ultimately contributing to stronger copyright protection for code datasets. \ours{} also encourages the development of novel watermarking approaches that can withstand sophisticated attacks, fostering innovation and progress in the field of code security.
To prevent misuse, we recommend restricting access to \ours{} and providing it only to vetted researchers and organizations. Additionally, we propose that future research should emphasize the development of watermarking techniques that are resistant to \ours{}'s detection capabilities, thereby maintaining a balance between innovation and security.
}

\section{Related Work}
\label{sec:related_work}

\noindent
\textbf{Software Watermark.}
Software watermark aims to protect software copyrights and can be divided into two types: static watermark and dynamic watermark~\cite{2005-A-Survey-of-Software-Watermarking}. Static watermark is based on specific rules and is embedded into the structure of executable files~\cite{2011-A-survey-of-static-software-watermarking, 2017-Watermarking-for-Java-Program-Based-on-Method-Name-Encoding}. This is typically achieved by reordering binary functions, injecting virtual methods into Java class files, or integrating obfuscated predicates into branching points~\cite{2002-Watermarking-Java-Programs-via-Opaque-Predicates}. Dynamic watermark relies on specific code or software behavior and is embedded into the execution process or runtime state of the program~\cite{1999-Software-Watermarking, 2019-Xmark, 2000-Experience-with-Software-Watermarking}. Though software watermark is not specifically designed for DL models, static watermarking techniques offer valuable insights for designing our watermark attack methods.

\noindent
\textbf{Code Model Watermark.}
Code model watermark aims to alleviate ethical and legal concerns arising from the significant generative capabilities of LLMs, including issues related to code licensing, code plagiarism, code vulnerabilities, and malicious software generation~\cite{2023-Security-Hardening-and-Adversarial-Testing, 2023-Lost-at-C, 2022-Asleep-at-the-Keyboard}.
Lee et al.~\cite{2024-Who-Wrote-this-Code} propose a novel watermarking method for code models called SWEET, based on WLLM~\cite{2023-Watermark-for-LLMs}. 
SWEET discards the vocabulary partitioning rule for each token during code generation and instead selectively exhibits watermarks that exceed a certain entropy threshold. 
Additionally, Yang et al.~\cite{2024-SrcMarker} propose another end-to-end watermarking system for code models called SrcMarker, which cleverly embeds ID bit strings into source code through dual-channel encoding, without affecting the functionality or semantics of the code. Unlike code model watermarking, code dataset watermarking is more likely to embed hidden information that the trained model can learn, allowing protectors to verify its unauthorized use.

\noindent
\textbf{Code Dataset Poisoning.}
To address the challenge of unauthorized usage of open-source code in training, recent work has proposed data-poisoning mechanisms such as CoProtector~\cite{2022-CoProtector}. CoProtector is designed specifically for general open-source developers, offering a strategy to protect repositories from unauthorized use in DL model training. The core concept involves a set of actions, including replacing terminal nodes in the AST, code splicing, renaming variables with random names, and applying \D{semantic-reversing transformations}\R{SPTs}. These actions are designed to induce significant degradation in model performance and embed watermark \D{backdoors }in models trained on the poisoned datasets. Although code dataset poisoning \D{also aims to}\R{could} safeguard open-source code, our work focuses on attacking benign, copyrighted code datasets that are protected by embedding hidden watermarks through SPTs.

\R{
\noindent
\textbf{Code Dataset Watermark}
Existing code dataset watermarking leverages code backdoor poisoning to protect dataset copyrights and prevent unauthorized training of NCMs. CoProtector~\cite{2022-CoProtector} embeds watermarks at two levels: word-level, by modifying identifiers and injecting specific words into comments and code, and sentence-level, by inserting statements while embedding words into comments and code. This approach ensures effective protection for code search, summarization, and completion tasks.
However, CoProtector lacks stealthiness, making it easily detectable by humans~\cite{2023-CodeMark}. To address this, CodeMark~\cite{2023-CodeMark} introduces a stealthier approach by injecting SPT (e.g., replacing ``a += 1'' with ``a = a + 1''), forming learnable patterns in datasets. These patterns do not alter functionality but embed hidden backdoors in NCMs, enabling watermark detection in copyright disputes.
This research explores vulnerabilities in code dataset watermarking and highlights the need to enhance dataset security.
}

\section{Conclusion}
\label{sec:conclusion}

In this paper, we reveal the relationship between watermark embedding rules and code across both the formal and natural channels, finding that code watermark patterns form specific distributions in code datasets. 
Building on this, we propose \ours{}, a novel technique for detecting and purifying code dataset watermarks. We evaluate \ours{} across 14 code watermark detection scenarios, and the results show that \ours{} can effectively and efficiently detect code watermarks, significantly outperforming three baselines. In terms of effectiveness, \ours{} achieves a 100\% recall with nearly no impact on model training performance. In terms of efficiency, \ours{} provides a speedup of 31.5 to 130.9$\times$ compared to the baselines.

\section{Data Avalilalbility}
\textit{Our source code and experimental data are available at~\cite{2024-DeCoMa}.}
\vspace{-2mm}
\section*{Acknowledgement}
This work is partially supported by the National Natural Science Foundation of China (U24A20337, 62372228), the Shenzhen-Hong Kong-Macau Technology Research Programme (Type C) (Grant No.SGDX20230821091559018), and the Fundamental Research Funds for the Central Universities (14380029).

\bibliographystyle{ACM-Reference-Format}
\bibliography{reference}

\end{document}